\documentclass[prd,onecolumn,preprintnumbers,floats,floatfix,apsrev,amsmath,amsfonts,nofootinbib,superscriptaddress]{revtex4-2}
\usepackage{color}
\usepackage{graphicx}
\usepackage{dcolumn}
\usepackage{natbib}
\usepackage{hyperref}
\usepackage{hypernat}
\usepackage{natbib}
\usepackage{soul}
\usepackage{amssymb}
\usepackage{orcidlink}
\usepackage{placeins}

\newcommand{\dd}{{\rm d}}

\newcommand{\p}{\partial}

\newcommand{\be}{\begin{equation}}
\newcommand{\en}{\end{equation}}

\newcommand{\paren}[1]{\left({#1}\right)}
\newcommand{\discont}[1]{\left[{#1}\right]_\Sigma}

\newcommand{\eff}{\text{eff}}

\newcommand{\vk}{\mathbf{k}}
\newcommand{\vx}{\mathbf{x}}
\newcommand{\vy}{\mathbf{y}}
\newcommand{\vz}{\mathbf{z}}
\newcommand{\vE}{\mathbf{E}}
\newcommand{\vB}{\mathbf{B}}
\newcommand{\vA}{\mathbf{A}}
\newcommand{\cE}{\mathcal{E}}
\newcommand{\cB}{\mathcal{B}}
\newcommand{\vU}{\mathbf{U}}
\newcommand{\vFc}{\boldsymbol{\mathcal{F}}}
\newcommand{\bL}{\bar{\mathcal{L}}}

\begin{document}
\title{Finite-amplitude wave dynamics in weak-field-equivalent nonlinear electrodynamics in $\left(2+1\right)$ dimensions}

\author{M.~L.~Pe\~{n}afiel\, \orcidlink{0000-0002-7700-1787}}
\email{miguelpenafiel@upb.edu}
\affiliation{FIA - Facultad de Ingeniería y Arquitectura, Universidad Privada Boliviana, Camino Achocalla Km 3.5, La Paz, Bolivia.}

\author{E.~Veneros\, \orcidlink{0009-0005-3667-3746}}
\email{erwinveneros@upb.edu}
\affiliation{FIA - Facultad de Ingeniería y Arquitectura, Universidad Privada Boliviana, Camino Achocalla Km 3.5, La Paz, Bolivia.}

\author{A.~Urquidi\, \orcidlink{0000-0001-7172-862X}}
\email{arielurquidi@upb.edu}
\affiliation{FIA - Facultad de Ingeniería y Arquitectura, Universidad Privada Boliviana, Camino Achocalla Km 3.5, La Paz, Bolivia.}

\begin{abstract}
We examine finite-amplitude wave propagation in $\paren{2+1}$-dimensional nonlinear electrodynamics using Born-Infeld, logarithmic, and quadratic polynomial models whose Lagrangians agree through quadratic order in the electromagnetic invariant. Despite this weak-field agreement, characteristic and linearized analyses on homogeneous magnetic background reveal distinct propagation speeds and hyperbolicity domains. In the weak-background limit, the first coefficient at which Lagrangians differ also controls the leading nonlinear variation of the simple-wave characteristic velocity. Exact right-moving simple-waves, corroborated by finite-volume evolution, show rigid Born-Infeld propagation but finite-time gradient catastrophes in the logarithmic and polynomial models, substantially earlier in the latter for matched pulses. These catastrophes can occur while the field remains finite and the system remains hyperbolic, showing that weak-field agreement does not determine finite-amplitude dynamics.
\end{abstract}

\maketitle

\section{Introduction} \label{sec:intro}

Maxwell electrodynamics is the archetype of a classical field theory, describing electromagnetic phenomena across a broad range of physical scales. In $\paren{2+1}$ dimensions, it still supports propagating waves, but the magnetic field is a pseudoscalar and a free wave has only one physical polarization \cite{Lapidus1982,Boito2020}. Experiments on atomically thin materials, including the direct imaging of propagating graphene plasmons, provide an observational motivation for studying planar wave phenomena  \cite{Fei2012,Chen2012}. On the theoretical side, studies of planar Maxwell radiation \cite{Boito2020} and of wave propagation under linear and nonlinear constitutive laws \cite{Goulart2022,Bittencourt2023} have clarified distinct aspects of planar propagation. The finite-amplitude evolution of waves in $\paren{2+1}$-dimensional nonlinear electrodynamics, however, remains comparatively unexplored.

Nonlinear extensions of electrodynamics (NLED) are often motivated by both foundational questions in classical electrodynamics and by effective descriptions of quantum vacuum polarization \cite{Born1934,Heisenberg1936,BialynickaBirula1970}. A particular feature of these theories is that the corresponding field equations can be cast in a form analogous to Maxwell's theory in material media, with nonlinear constitutive laws that depend on the electromagnetic field itself. NLED theories can exhibit exceptional propagation or the steepening of smooth waves \cite{Boillat1970,Plebanski1970,Novello2000,Heyl1998,Kadlecova2019}. In a lower dimensional setting, radiation under a conformally invariant NLED has been studied \cite{Cataldo2014}, as well as wave propagation under uniform fields in $d\ge2$ spatial dimensions \cite{Soares2024}.

A model of NLED intended to extend Maxwell electrodynamics should reproduce its weak-field predictions, while nonlinear effects become appreciable at field strengths set by the model's characteristic scale. Further tests of physical admissibility include causal propagation, hyperbolicity, positivity conditions related to perturbative unitarity, invertibility of the constitutive map, and energy conditions \cite{Shabad2011,Abalos2015,Schellstede2016,Bronnikov2001,Russo2024}. These criteria can hold over a common field range without fixing the finite-amplitude dynamics of theories whose weak-field Lagrangians agree through quadratic order in the electromagnetic invariant $F$. We examine this possibility for Born-Infeld \cite{Born1934,Ellis2017}, logarithmic \cite{Gaete2014}, and a quadratic polynomial model by comparing simple-wave propagation on homogeneous magnetic backgrounds.

Finite-amplitude wave evolution is an initial-value problem for the nonlinear field equations. In this work, we compare the three aforementioned NLED models by constructing exact characteristic solutions for right-moving simple-waves and evolving each corresponding initial state numerically. The characteristic map determines when an initially smooth profile develops a divergent spatial gradient. Together, these approaches show rigid propagation along the Born-Infeld simple-wave and finite-time steepening in the logarithmic and polynomial models, as well as different amplitude ranges over which the waves remain hyperbolic.  Moreover, in the weak-background regime we show analytically that the first higher-order coefficient distinguishing the matched Lagrangians determines their leading difference in simple-wave steepening.

The paper is organized as follows. In Sec. \ref{sec:2} we introduce Maxwell and nonlinear electrodynamics in $\paren{2+1}$ dimensions, and in Sec. \ref{sec:3} we define the weak-field matched models to be compared. In Sec. \ref{sec:4} we study propagation on a homogeneous magnetic background using Hadamard's discontinuity method and a weak-amplitude perturbative analysis. In Sec. \ref{sec:5} we examine the physical admissibility of the models, while in Sec. \ref{sec:6} we present the finite-amplitude characteristic and numerical analyses. We summarize our conclusions and future directions in Sec. \ref{sec:concl}. Unless otherwise stated, we use natural units where $\hbar=c=1$ and the Minkowski metric in a $(2+1)$-dimensional spacetime adopts the signature $\eta_{ab}=\text{diag}\paren{1,-1,-1}$.

\section{Electrodynamics in $(2+1)$ dimensions} \label{sec:2}

\subsection{Maxwell electrodynamics in $(2+1)$ dimensions}
Let us first summarize the distinctive features of Maxwell electrodynamics once the spacetime under consideration possesses only 3 dimensions.
The Faraday tensor is a totally antisymmetric rank 2 tensor constructed from the derivatives of the electromagnetic potential vector $A_a$ which takes the form $F_{ab}=\p_a A_b-\p_b A_a$. Thus, given the dimensionality of the spacetime under inspection, the Faraday tensor possesses only three independent components which are identified with the two electric field components and a single magnetic component according to
\be
E_i=F_{i0}\quad , \quad B=F_{12}\ .
\en
Thus, the electric field in $(2+1)$ dimensions remains a spatial vector, while the magnetic field possesses only one independent component and transforms as a pseudoscalar. From the Faraday tensor, one can construct the single Lorentz invariant 
\be \label{eq:invF}
F=-\frac{1}{4}F_{ab}F^{ab}=\frac{1}{2}\paren{E^2-B^2}\ .
\en
Maxwell electrodynamics follows from the action
\be
S_\text{Max}=\int \dd^3 x F\ ,
\en
and the corresponding source-free field equations are
\begin{subequations}
\begin{align}
\p_a F^{ab}&=0\ ,\\
\p_{[a} F_{bc]}&=0\  , \label{eq:Bianchi}
\end{align}
\end{subequations}
where the last equation corresponds to the Bianchi identity. In terms of the electric and magnetic fields, the latter equations can be written in their planar form as
\begin{align}
\p_x E_x+\p_y E_y&=0\ , \\
\p_t E_x-\p_y B&=0\ , \\
\p_t E_y+\p_x B&=0\ , \\
\p_x E_y-\p_y E_x+\p_t B&=0\ .
\end{align}

Equivalently, taking the derivative of the Bianchi identity in Eq. \eqref{eq:Bianchi} results in
\be
\p^c\p_c F_{ab}=0\ .
\en
Thus, each independent component of the electromagnetic tensor satisfies a wave equation. For a plane-wave configuration of the form $F_{ab}\propto \exp\paren{-i\omega t+i\vk\cdot\vx},$ the wave equation yields the dispersion relation $\omega^2=k^2$. Furthermore, from Gauss' law and the Bianchi identity we obtain
\be
\vk\cdot\vE=0 \ , \quad B=\frac{\epsilon^{ij}k_iE_j}{\omega}
\en
Hence there is only one independent transverse polarization in $(2+1)$ dimensions. In particular, for a null Maxwell plane wave one has $B^2=E^2$, thus $F=0$.

Finally, the energy-momentum tensor of Maxwell's electrodynamics is given by 
\be \label{eq:TabMax}
T_{ab}=-F_{ac}F_b^{\ c}+\frac{1}{4}\eta_{ab}F_{cd}F^{cd}\ ,
\en
for which the trace is $T=T^{a}_{\ a}=F$, which is generically nonvanishing. Thus, unlike the four-dimensional case, the standard formulation of Maxwell electrodynamics in three spacetime dimensions does not possess manifest Weyl invariance.\footnote{However, in three spacetime dimensions it is possible to construct a local description of the theory using a scalar gauge field, which gives rise to an improved, traceless, energy-momentum tensor \cite{Deser1994}.} 

\subsection{Nonlinear electrodynamics in $(2+1)$ dimensions}
In $(2+1)$ dimensions it is possible to construct only one Lorentz invariant, namely $F=-F_{ab}F^{ab}/4$. We shall consider nonlinear extensions of electromagnetism for which Lagrangians are functions of this invariant, i.e., $\mathcal{L}=\mathcal{L}\paren{F}$. Defining $\mathcal{L}_F\equiv\p\mathcal{L}/\p F$, the field equations stemming from such a Lagrangian are
\be \label{eq:dyneqNLED}
\p_a\paren{\mathcal{L}_F F^{ab}}=0\ ,
\en
while the Bianchi identity remains unchanged.

It is convenient to introduce the electric displacement and magnetic excitation field through the constitutive relations
\be
D_i\equiv\frac{\p\mathcal{L}}{\p E_i}=\mathcal{L}_FE_i\ ,
\en
and 
\be
H\equiv-\frac{\p\mathcal{L}}{\p B}=\mathcal{L}_F B\ .
\en
In terms of these quantities, the nonlinear field equations read
\begin{align} \label{eq:NLEDfieldeq1}
\p_x D_x+\p_y D_y&=0\ , \\
\p_t D_x-\p_y H&=0\ , \\
\p_t D_y+\p_x H&=0\ ,\\
\p_t B+\p_xE_y-\p_y E_x&=0\ . \label{eq:NLEDfieldeq4}
\end{align}

Since $\mathcal{L}_F$ is itself a function of the electromagnetic field, we may expand Eq. \eqref{eq:dyneqNLED} as
\be \label{eq:fieldNLEDexpanded}
\mathcal{L}_F\p_a F^{ab}+\mathcal{L}_{FF}F^{ab}\p_a F=0\ ,
\en
where we have defined $\mathcal{L}_{FF}\equiv\p^2\mathcal{L}/\p F^2$. Thus, Eq. \eqref{eq:fieldNLEDexpanded} constitutes a quasilinear system whose characteristic structure depends on the electromagnetic background.

The energy-momentum tensor associated with a generic $\mathcal{L}\paren{F}$ NLED is given by 
\be\label{eq:Tab}
T_{ab}=-\mathcal{L}_FF_{ac}F_b^{\ c}-\eta_{ab}\mathcal{L}\paren{F}\ .
\en
Straightforward inspection shows that theories possessing a manifest conformal invariance need to have the dependence $\mathcal{L}\paren{F}\propto\vert F\vert^{3/4}$ and Maxwell electrodynamics does not belong to this class of theories. Nonetheless, we shall consider theories which recover Maxwell electrodynamics as a weak-field approximation and are thus not manifestly conformally invariant. 

\section{Weak-field-equivalent nonlinear electrodynamics} \label{sec:3}

We are interested in analyzing models of electrodynamics that are not only equivalent to Maxwell electrodynamics in the weak-field limit, but equivalent to each other at the second perturbative order. This will allow us to study at what extent theories that are indistinguishable at sufficiently small field strengths can lead to different propagation properties once nonlinear effects become relevant. 

We shall consider three models of NLED: namely Born-Infeld (BI), logarithmic, and polynomial electrodynamics whose Lagrangians are written as

\begin{align}
\mathcal{L}_\text{BI}&=b_\text{BI}^2\paren{1-\sqrt{1-\frac{2 F}{b_\text{BI}^2}}}\ , \\
\mathcal{L}_\text{log}&=-b_\text{log}^2\ln\paren{1-\frac{F}{b_\text{log}^2}}\ , \label{eq:Llog}\\
\mathcal{L}_\text{p}&=F+\frac{F^2}{2b_\text{p}^2}\ , \label{eq:Lpoly}
\end{align}
respectively, and the constants $b_\text{BI}$, $b_\text{log}$ and $b_\text{p}$ can be thought of as nonlinearity scales of each theory. Hence, in a sense, these constants fix the nonlinearity scale of each theory and determine the field scale at which nonlinear departures from Maxwell electrodynamics become apparent. 

Notice that as $F\to0$, all three models introduced above recover Maxwell electrodynamics as the leading order contribution. In what follows, we shall impose a stronger matching condition, requiring the theories to agree not only in their Maxwellian limit but also in their leading nonlinear correction. Expanding the models above around $\vert F\vert/b^2\ll1$, we obtain
\begin{align}
\mathcal{L}_{\text{BI}}&\approx F+\frac{F^2}{2b_\text{BI}^2}+\frac{F^3}{2b_\text{BI}^4}+\mathcal{O}\paren{F^4/b_\text{BI}^6}\ , \\
\mathcal{L}_\text{log}&\approx F+\frac{F^2}{2b_\text{log}^2}+\frac{F^3}{3b_\text{log}^4}+\mathcal{O}\paren{F^4/b_\text{log}^6}\ ,
\end{align}
while the polynomial model is, by construction, 
\be
\mathcal{L}_p=F+\frac{F^2}{2b_\text{p}^2}\ .
\en

We now require the three nonlinear theories to possess the same leading order correction to Maxwell electrodynamics. Hence, direct comparison of the quadratic terms in the above expansions yield
\be
b_\text{BI}=b_\text{log}=b_{p}\equiv b\ ,
\en
where $b$ now defines a common nonlinear field scale. Hence, we can write the expansions for the three second-order equivalent theories as
\be \label{eq:Lgen}
\mathcal{L}^{(i)}\paren{F}=F+\frac{F^2}{2b^2}+c_i\frac{F^3}{b^4}+\mathcal{O}\paren{F^4/b^6}\ ,
\en
where $c_i$ correspond to the cubic term coefficient for each theory, namely
\[
c_\text{BI}=\frac{1}{2}\ , \quad c_\text{log}=\frac{1}{3}\ , \quad c_\text{p}=0\ .
\]
In this sense, the models considered here form a class of weak-field-equivalent nonlinear electrodynamics: they possess both the same Maxwellian limit and the same leading nonlinear correction, while differing in their higher-order completion. Consequently, any observable whose weak-field behavior depends only on the terms up to $F^2$ will be unable to distinguish between these theories. Their differences can first emerge only through quantities sensitive to the higher-order structure of the corresponding Lagrangians.

At first sight, one might expect that, because the three models agree through quadratic order in $F$, their predictions for physical observables would also remain equivalent to the same perturbative order. However, observables involving derivatives of the Lagrangian can distinguish the theories at lower powers of the background field. From Eq. \eqref{eq:Lgen}, we have
\be
\mathcal{L}^{(i)}_F=1+\frac{F}{b^2}+3c_i\frac{F^2}{b^4}+\mathcal{O}\paren{F^3/b^6}\ ,
\en
and
\be
\mathcal{L}^{(i)}_{FF}=\frac{1}{b^2}+6c_i\frac{F}{b^4}+\mathcal{O}\paren{F^2/b^6}\ .
\en

Particularly, when considering wave propagation in theories of NLED we have Eq. \eqref{eq:fieldNLEDexpanded}, which involves second order derivatives of the Lagrangian. Consequently, two theories that are weak-field equivalent up to second order need not remain equivalent at the same order when considering their propagation properties. This distinction motivates two notions of weak-field equivalence at the level of a specified observable: first, we can consider equivalent theories at a given perturbative order of the Lagrangian, and next consider equivalent theories with respect to concrete physical observables.

\section{Wave propagation in a homogeneous background} \label{sec:4}

\subsection{Effective geometry}
Let us start by inspecting the propagation of electromagnetic disturbances from the characteristic structure of the field equations. We shall employ the Hadamard method of field discontinuities. Let $\Sigma$ denote a characteristic hypersurface describing the propagation of a wavefront, with a normal covector given by
\be
k_a=\p_a \Sigma\ .
\en
The Hadamard method is based on assessing the field discontinuities across the hypersurface. Let us denote the discontinuity across $\Sigma$ by square brackets, we assume that the field is continuous across $\Sigma$, while the first derivatives of the field exhibit a finite discontinuity, i.e.,
\be \label{eq:discontinuities}
\discont{F_{ab}}=0\ ,\quad \discont{\p_c F_{ab}}=k_c f_{ab}\ ,
\en
where $f_{ab}=-f_{ba}$ represents the discontinuity amplitude.

Applying the Hadamard method in Eq. \eqref{eq:discontinuities} to the invariant $F$ we find
\be
\discont{\p_c F}=-\frac{1}{2}\zeta k_c\ ,
\en
where we have defined $\zeta\equiv F^{ab}f_{ab}$. Furthermore, applying this to the field equation \eqref{eq:fieldNLEDexpanded} we obtain 
\be \label{eq:dField}
\mathcal{L}_Fk_a f^{ab}-\frac{1}{2}\mathcal{L}_{FF}\zeta F^{ab}k_a=0\ ,
\en
while taking the discontinuity of the Bianchi identity \eqref{eq:Bianchi} leads to 
\be \label{eq:dBianchi}
k_a f_{bc}+k_b f_{ca}+k_c f_{ab}=0\ .
\en
The latter equations constitute the algebraic system governing the propagation of field discontinuities. Contracting Eq. \eqref{eq:dBianchi} with $k^a$ such that
\[
k^2 f_{bc}+k_bk^af_{ca}+k_ck^af_{ab}=0\ , 
\]
where $k^2\equiv\eta^{ab}k_ak_b$. We can further multiply the latter equation by $F^{bc}$, and use Eq. \eqref{eq:dField} to obtain
\be \label{eq:37}
\zeta\paren{\eta^{ab}-\frac{\mathcal{L}_{FF}}{\mathcal{L}_F}F^{ac}F^{b}_{\ c}}k_ak_b=0\ .
\en
The factor $\zeta=0$ in Eq. \eqref{eq:37} does not by itself establish an additional characteristic branch. As shown in Appendix \ref{ap:A}, the full discontinuity equations admit no nontrivial $\zeta=0$ solution on a nonzero homogeneous magnetic background. For the generic branch $\zeta\ne0$, the characteristic surfaces are therefore determined by 
\be\label{eq:geff}
g_{\eff}^{ab}k_ak_b=0\ ,
\en
where the effective contravariant metric is defined as
\be \label{eq:geff1}
g^{ab}_\eff=\eta^{ab}-\frac{\mathcal{L}_{FF}}{\mathcal{L}_F}F^{ac}F^b_{\ c}\ .
\en
Equation \eqref{eq:geff} shows that high-frequency electromagnetic disturbances propagate along null directions of the field-dependent effective metric. 

In NLED in $\paren{3+1}$ dimensions, the characteristic structure may depend on the polarization, leading to vacuum birefringence; Born-Infeld electrodynamics is a notable exception. In $\paren{2+1}$ dimensions, however, only one physical polarization is present, and its propagation is governed by Eq. \eqref{eq:geff}.

We now consider the case of wave propagation under a homogeneous magnetic configuration. Let the external field be denoted by an over-bar and the homogeneous magnetic field under inspection be $B_0$, i.e., the background fields are
\be
\bar{E}_x=\bar{E}_y=0\ , \quad \bar{B}=B_0=\text{const.}
\en
Thus, direct substitution into Eq. \eqref{eq:geff1} gives the effective metric
\be \label{eq:geffB0}
g_\eff^{ab}=\text{diag}\left[\mathcal{L}_F,-\mathcal{L}_F+B_0^2\mathcal{L}_{FF},-\mathcal{L}_F+B_0^2\mathcal{L}_{FF}\right]\ ,
\en
where all derivatives are evaluated at the background configuration, unless otherwise stated. Let us take the characteristic covector, $k_a$, to be given by
\be
k_a=\paren{-\omega,k_x,k_y}\ ,
\en
Plugging this into the propagation condition, Eq. \eqref{eq:geff}, and writing the dispersion relation as $\omega^2=v^2 k^2$, leads the characteristic velocity of propagation to be
\be \label{eq:vchar}
v^2=1-B_0^2\frac{\mathcal{L}_{FF}}{\mathcal{L}_F}\ .
\en
Real characteristic propagation requires $v^2>0$; this requirement is linked to the hyperbolicity of the characteristic equation. Moreover, demanding that the effective characteristic cone does not extend outside the Minkowski light cone imposes $v^2\leq1$; a condition associated with causal propagation of the theory under inspection. These conditions will be analyzed in more detail in Sec.~\ref{sec:5}; however, here they will be used only to identify the physically relevant magnetic branch of each model.

To compare the theories introduced in Sec.~\ref{sec:3}, it is convenient to define the dimensionless background strength
\begin{equation}
    \beta^2\equiv\frac{B_0^2}{b^2}.
    \label{eq:xmagnetic}
\end{equation}
For Maxwell electrodynamics, $\mathcal{L}_F=1$ and $\mathcal{L}_{FF}=0$, so that
\be\label{eq:vMax}
    v_{\rm M}^2=1\ .
\en
For Born-Infeld electrodynamics,
\begin{equation}
    \mathcal{L}_F=\left(1-\frac{2F}{b^2}\right)^{-1/2},\qquad  \mathcal{L}_{FF}=\frac{1}{b^2}\left(1-\frac{2F}{b^2}\right)^{-3/2},
\end{equation}
and the characteristic velocity in Eq. \eqref{eq:vchar} gives
\be \label{eq:v2BI}
v_\text{BI}^2=\frac{1}{1+\beta^2}\ .
\en
Thus, we have $v^2_\text{BI}\le1$ for every value of $\beta$ \cite{Gibbons2001}.

Analogously, for logarithmic electrodynamics we have,
\be
\mathcal{L}_F=\frac{1}{1-F/b^2}\ ,\qquad \mathcal{L}_{FF}=\frac{1}{b^2(1-F/b^2)^2},
\en
leading to
\be  \label{eq:v2log}
v_{\log}^2=\frac{1-\beta^2/2}{1+\beta^2/2}\ .
\en
The denominator remains positive in the magnetic sector, while real propagation requires
\begin{equation}    \label{eq:logRange}
    \beta^2<2\ .
\end{equation}
At $\beta^2=2$, the characteristic velocity vanishes and the effective metric becomes degenerate.

Finally, for the polynomial model,
\be
    \mathcal{L}_F=1+\frac{F}{b^2}\ ,\qquad \mathcal{L}_{FF}= \frac{1}{b^2},
\en
and therefore
\be \label{eq:v2poly}
 v_\text{p}^2 =\frac{1-3\beta^2/2}{1-\beta^2/2}\ .
\en
On the branch continuously connected to Maxwell electrodynamics, the simultaneous requirements of hyperbolic and subluminal propagation give
\begin{equation}    \label{eq:PolyRange}
    \beta^2<\frac{2}{3}\ .
\end{equation}
The effective metric becomes degenerate at $\beta^2=2/3$.

\begin{figure}[t]
\centering
\includegraphics[width=0.47\textwidth]{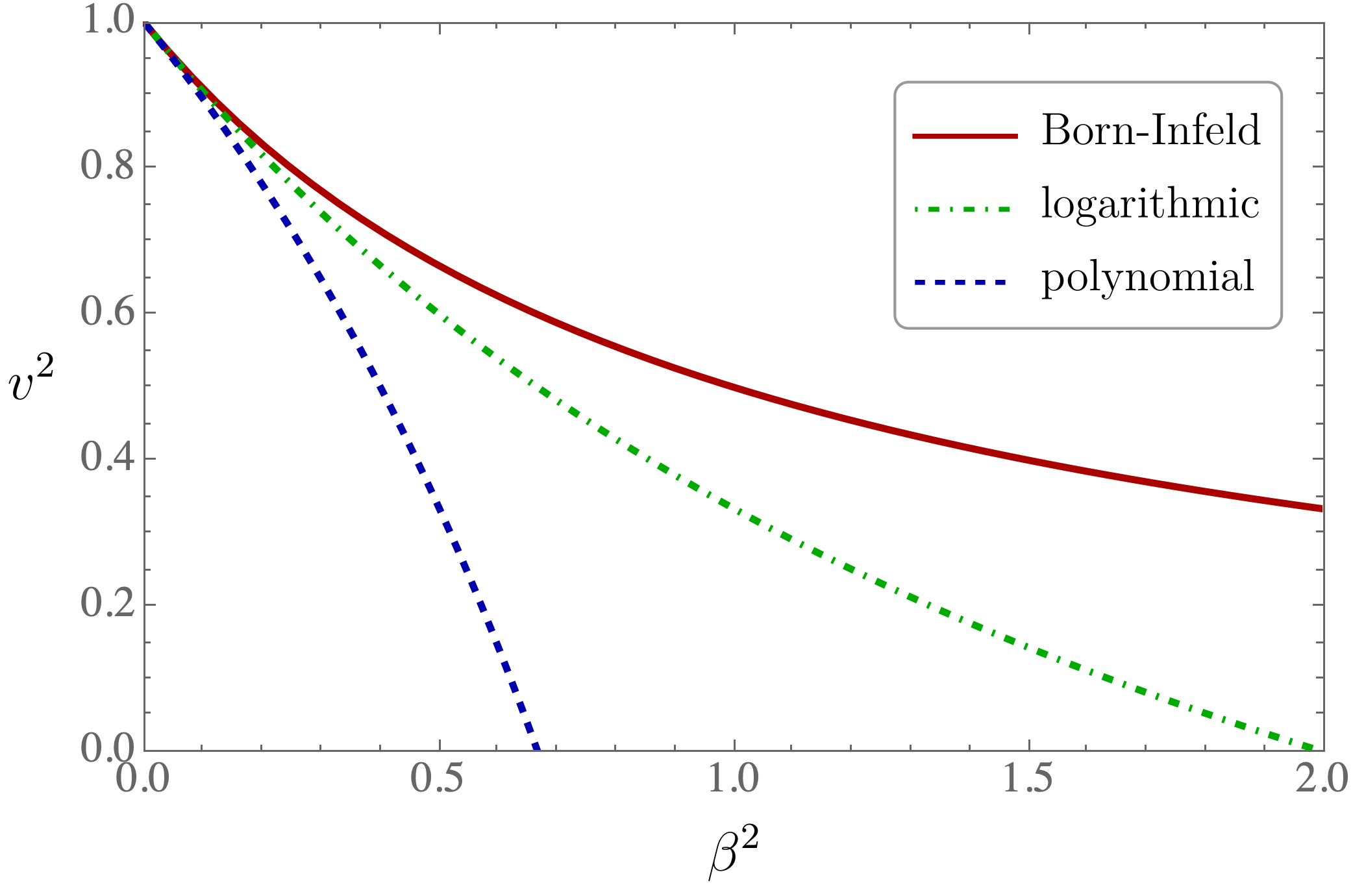}
\caption{Squared characteristic velocity, $v^2$, versus the dimensionless magnetic background, $\beta^2=B_0^2/b^2$ for Born-Infeld, logarithmic, and polynomial electrodynamics. In the common hyperbolic range $0<\beta^2<2/3$, $v_{\text{BI}}^2>v_{\log}^2>v_p^2$.}
\label{fig:fig1}
\end{figure}

We can inspect the weak-field regime for the propagation velocities for each theory. Expanding Eqs. \eqref{eq:v2BI}, \eqref{eq:v2log}, and \eqref{eq:v2poly} around $\beta\ll1$ we obtain
\begin{align}
v^2_\text{BI}&=1-\beta^2+\beta^4+\mathcal{O}\paren{\beta^6}\ , \\
v^2_{\log}&=1-\beta^2+\frac{1}{2}\beta^4+\mathcal{O}\paren{\beta^6}\ , \\
v^2_p&=1-\beta^2-\frac{1}{2}\beta^4+\mathcal{O}\paren{\beta^6}\ .
\end{align}
Hence the three theories become distinguishable at $\mathcal{O}\paren{\beta^4}$. Moreover, within the common hyperbolic domain for the three theories, $0<\beta^2<2/3$, we obtain
\be
v_\text{BI}^2-v_{\log}^2=\frac{\beta^4}{\paren{1+\beta^2}\paren{2+\beta^2}}>0\ ,
\en
and
\be
v_{\log}^2-v_p^2=\frac{4\beta^4}{\paren{2-\beta^2}\paren{2+\beta^2}}>0\ .
\en
Hence over the common hyperbolic domain for the three theories we have the hierarchy
\be
v_\text{BI}>v_{\log}>v_p\ .
\en
Thus, as the background magnetic field intensity increases, a separation among the effective light cones for each of the weak-field-equivalent theories begins. Using the latter equations for the characteristic velocities for each theory, we can evaluate the expressions over the hyperbolicity domain of each theory. Fig. \ref{fig:fig1} displays the characteristic velocities of the three nonlinear theories and verifies that wave propagation within Born-Infeld electrodynamics remains hyperbolic throughout the magnetic sector, while logarithmic and polynomial models possess finite limiting background strengths that are linked to the lack of hyperbolicity beyond these critical values.   

\subsection{Linear perturbations around the magnetic background}

Let us now study the small-amplitude perturbations around a given homogeneous magnetic background. This approach, in contrast to the Hadamard construction above, does not rely on a high-frequency approximation, but instead, it is independent of the wave frequency. We shall consider the field intensities of the wave to be small in comparison with the background magnetic field of magnitude $B_0$. Thus, the perturbations take the form
\be \label{eq:pertset}
E_i=e_i\ , \quad B=B_0+b\ ,
\en
where $e_i,b\ll B_0$. Therefore, the electromagnetic invariant F can be expressed as $F=\bar{F}+\delta F$, where $\bar{F}=-B_0^2/2$ is the invariant corresponding to the background field, while, $\delta F=-B_0 b$ is a first order perturbation. In the following, all quantities expressed with an overbar will be evaluated at the background field configuration. Hence, expanding $\mathcal{L}_F\paren{F}$ around the background gives
\be
\mathcal{L}_F=\bL_F+\bL_{FF}\delta F+\mathcal{O}\paren{2}\ .
\en
The electric displacement therefore reduces to 
\be \label{eq:deltaD}
\delta D_i=\bL_F e_i\ , 
\en
while the magnetic excitation takes the form
\be \label{eq:deltaH}
\delta H=\paren{\bL_F-\bL_{FF}B_0^2}b\ .
\en

Substitution of Eqs. \eqref{eq:deltaD} and \eqref{eq:deltaH} into the system of Eqs. \eqref{eq:NLEDfieldeq1} - \eqref{eq:NLEDfieldeq4} yields the linearized system
\begin{align}
\p_xe_x+\p_ye_y&=0\ , \label{eq:linearGauss}\\
\bL_F \p_t e_x-\paren{\bL_F-\bL_{FF}B_0^2}\p_y b&=0\ ,\\
\bL_F\p_t e_y+\paren{\bL_F-\bL_{FF}B_0^2}\p_x b&=0\ , \label{eq:ycomp1}\\
\p_t b+\p_xe_y-\p_ye_x&=0 \ . \label{eq:linearFaraday}
\end{align}
Equations \eqref{eq:linearGauss}--\eqref{eq:linearFaraday} describe the complete first-order dynamics of electromagnetic disturbances around the homogeneous magnetic background. Since $\bL_F$ and $\bL_{FF}$ are constants for the background under consideration, the above system has constant coefficients. 

Let us now seek plane wave solutions for the linearized system of the form $\paren{e_i,b}\propto\exp\left[i\paren{\vk\cdot\vx-\omega t}\right]$. Substituting the \emph{ansatz} into Eq. \eqref{eq:linearGauss} gives
\be
\vk\cdot\mathbf{e}=0\ ,
\en
thus the electric perturbation is transverse to the direction of propagation. On the other hand, plugging the \emph{ansatz} into Eq. \eqref{eq:linearFaraday} results in 
\be \label{eq:bfieldex}
b=\frac{\epsilon^{ij}k_i}{\omega} e_j\ .
\en
Hence the magnetic perturbation is completely determined by the transverse electric component and does not constitute an independent degree of freedom, i.e., when considering electromagnetic wave propagation in a  $(2+1)$ dimensional spacetime, only one physical polarization exists and birefringence therefore cannot occur.  

Plugging the \emph{ansatz} into Eq. \eqref{eq:ycomp1} and using Eq. \eqref{eq:bfieldex} leads to the dispersion relation
\be \label{eq:disprelNLED}
\omega^2=\paren{1-B_0^2\frac{\bL_{FF}}{\bL_F}}k^2\ .
\en
Hence, the velocity of propagation of the perturbation coincides with the characteristic velocity obtained from the effective metric (c.f. Eq. \eqref{eq:vchar}). 

The Hadamard construction determines the local characteristic cone governing the propagation of weak discontinuities and is closely related to the geometric-optics description. By contrast, the perturbative analysis above assumes only that the disturbance amplitude is small relative to the background and imposes no restriction on its frequency. Nevertheless, both approaches yield the same propagation velocity. Unlike the geometric-optics description, however, the linearized calculation is not restricted to the high-frequency regime.

Let us stress that the phase and group velocities obtained from the dispersion relation are equal and no chromatic dispersion appears when considering $\mathcal{L}\paren{F}$ theories. Thus, scanning the frequency domain within the linearized theory cannot reveal additional dispersive effects. Any further distinction among the models must instead arise from the full finite-amplitude nonlinear dynamics.

\section{Physical admissibility of the nonlinear models} \label{sec:5}

\subsection{Perturbative stability, hyperbolicity and causality} \label{ssec:5.1}

The propagation analysis of the previous section revealed that the characteristic structure of the theory is governed by the combinations $\mathcal{L}_F$ and $\mathcal{L}_F+2 F\mathcal{L}_{FF}$. The physical significance of such derivatives of the NLED Lagrangian can also be understood from the dynamics of small fluctuations around a background field, where the quadratic action determines the kinetic and constitutive response of the propagating degrees of freedom \cite{Shabad2011,Neves2021,PerezGarcia2023}. We shall apply this construction to the $(2+1)$ dimensional cases considered here, and use it to identify the parameter regions in which the NLED models define a physically admissible evolution.

By considering the perturbative setting of Eq. \eqref{eq:pertset}, the electromagnetic invariant can be expanded as
\be
F=\bar{F}+\delta F^{(1)}+\delta F^{(2)}+\mathcal{O}\paren{3}\ , 
\en
with 
\be
\delta F^{(1)}=-B_0b\ , \quad \delta F^{(2)}=\frac{1}{2}\paren{e^2-b^2}\ .
\en
Thus, expanding the Lagrangian around the background gives
\be
\mathcal{L}(F)=\bL+\bL_F\delta F^{(1)}+\bL_F\delta F^{(2)}+\frac{1}{2}\bL_{FF}\paren{\delta F^{(1)}}^2+\mathcal{O}\paren{3}\ .
\en
The linear term in the perturbation can be proved to be a total divergence of the action \cite{Karbstein2015,Neves2021,PerezGarcia2023}\footnote{Considering that the electromagnetic tensor can be written in terms of the background field and the perturbation, $F_{ab}=\bar{F}_{ab}+f_{ab}$, we have $b=f_{12}$ and 
\[
\mathcal{L}^{(1)}=\p_x\paren{-\bL_F B_0 a_2}+\p_y\paren{\bL_F B_0a_1}\ ,
\]
which constitutes a boundary term for the action.
}, and thus the first non-trivial contribution corresponds to 
\be \label{eq:quadL}
\mathcal{L}^{(2)}=\frac{1}{2}\bL_F e^2-\frac{1}{2}\paren{\bL_F+2\bar{F}\bL_{FF}}b^2\ .
\en
Equation \eqref{eq:quadL} provides a direct physical interpretation of both coefficients. The electric perturbation contains the time derivatives of the spatial gauge potential, while the magnetic perturbation contains only spatial derivatives. Hence, both terms are associated with a kinetic and a gradient contribution, respectively. Since we are interested in recovering Maxwell electrodynamics in the weak-field limit, we need to select the branch for which $\mathcal{L}_F\to1$ as $F\to0$, and the positivity of the first coefficient requires
\be \label{eq:cond1}
\mathcal{L}_F>0\ .
\en
On the other hand, demanding the positivity of the gradient term leads to
\be \label{eq:cond2}
\mathcal{L}_F+2F\mathcal{L}_{FF}>0\ .
\en
Hence, conditions \eqref{eq:cond1} and \eqref{eq:cond2} ensure that both the kinetic and gradient terms are stable in the linearized theory. 

These same conditions arise independently by analyzing the effective geometry. We begin by requiring the effective metric in Eq. \eqref{eq:geffB0} to be nondegenerate and to possess Lorentzian signature. Hence, demanding that $\mathcal{L}_F>0$ for us to recover Maxwell electrodynamics in the correct limit, we obtain 
\be \label{eq:hyper}
1+2F\frac{\mathcal{L}_{FF}}{\mathcal{L}_{F}}>0\ ,
\en
which is equivalent to requiring real characteristic propagation, $v^2>0$, as discussed in the previous section. 

Next, requiring that the characteristic light-cone for the NLED under consideration not extend outside the Minkowski light-cone implies the causality condition $v^2\le1$. Using Eq. \eqref{eq:hyper} and considering that for the purely magnetic sector, $F<0$, and $\mathcal{L}_F>0$, we have the causality condition
\be \label{eq:causal}
\mathcal{L}_{FF}\ge0\ .
\en

Therefore, the perturbative stability analysis, and the requirements of hyperbolicity and causality for the propagation lead to conditions \eqref{eq:cond1}, \eqref{eq:cond2} and \eqref{eq:causal}, respectively.  We shall now see that similar conditions are necessary for satisfying other physical requirements.

\subsection{Legendre map and constitutive degeneracy}

Nonlinear electrodynamics relates the electromagnetic fields $\paren{E_i,B}$ to the excitations $\paren{D_i,H}$ through field-dependent constitutive relations. It is therefore useful to ask when this map admits a local inverse \cite{SalazarI.1987}. Let us define the excitation tensor conjugate to $F_{ab}$ by
\be
E^{ab}\equiv-2\frac{\p\mathcal{L}}{\p F_{ab}}=\mathcal{L}_F F^{ab}\ .
\en
Its invariant, defined with the same convention as $F$ in Eq.~\eqref{eq:invF}, is
\be \label{eq:invP}
P\equiv-\frac{1}{4}E^{ab}E_{ab}=F\mathcal{L}_F^2\ .
\en
The map $F\mapsto P\paren{F}$ relates the Lagrangian and Hamiltonian-like formulations of the theory. Differentiation gives
\be \label{eq:dPdF}
\frac{dP}{dF}=\mathcal{L}_F\mathcal{C},\qquad
\mathcal{C}\equiv\mathcal{L}_F+2F\mathcal{L}_{FF}\ .
\en
On the homogeneous magnetic background, Eq.~\eqref{eq:vchar} implies $dP/dF=\mathcal{L}_F^2v^2$. Thus, within a connected Maxwell branch with $\mathcal{L}_F>0$ and $v^2>0$, the map is monotonic and locally invertible. At $\mathcal{C}=0$, both this derivative and the magnetic-background characteristic speed vanish. The inverse constitutive relations are derived in Appendix~\ref{ap:admissibility}; related constitutive branching has also been studied in Refs.~\cite{Bronnikov2001,Bronnikov2018}.

\subsection{Energy conditions} \label{sec:enCond}

For a homogeneous magnetic background, the stress tensor in Eq.~\eqref{eq:Tab} has the perfect-fluid form $T^a{}_{b}=\operatorname{diag}(\rho,-p,-p)$, with
\be
\rho=-\mathcal{L},\qquad p=\mathcal{L}-2F\mathcal{L}_F\ .
\en
The null, weak, dominant, and strong energy conditions then require, respectively, $\rho+p\ge0$; $\rho\ge0$ and $\rho+p\ge0$; $\rho\ge |p|$; and $\rho+p\ge0$ together with $p\ge0$. The last form is specific to $\paren{2+1}$ dimensions. Since $F<0$ on a nonzero magnetic background, $\rho+p=-2F\mathcal{L}_F\ge0$ on the Maxwell-connected branch. Direct evaluation for the three models shows that all four conditions hold throughout their respective hyperbolic magnetic-background domains. The detailed inequalities and model-specific energy densities and pressures are given in Appendix~\ref{ap:admissibility}.

The energy conditions impose no tighter restriction than hyperbolicity and causality in the magnetic sectors compared here. We therefore use those bounds to define the admissible homogeneous magnetic backgrounds for comparing finite-amplitude wave propagation. Hyperbolicity along the simple-wave profiles is assessed separately in Sec.~\ref{sec:6}.

\begin{table*}[t]
\centering
\begin{tabular}{c|c|c|c|c|c|c}
\hline\hline
Model & $\mathcal{L}_F>0$ & $\mathcal{C}>0$ & $\mathcal{L}_{FF}\ge0$ & $dP/dF>0$ & EC  & Admissible domain \\ \hline
  Born-Infeld    &     $x\ge0$              &      $x\ge0$           &      $x\ge0$           &    $x\ge0$       & Yes &   $0\le x<\infty$                \\ 
   Logarithmic   &          $x\ge0$         &         $x<2$        &        $x\ge0$          &      $x<2$     & Yes &             $0\le x<2$      \\ 
    Polynomial  &             $x<2$      &             $x<2/3$    &          $x\ge0$         &      $x<2/3$     & Yes &           $0\le x<2/3$        \\ \hline
      \hline
\end{tabular}
\caption{Admissibility conditions on homogeneous magnetic backgrounds, with $x=\beta^2$ and $\mathcal{C}=\mathcal{L}_F+2F\mathcal{L}_{FF}$. EC denotes all four energy conditions within the listed domain.}
\label{tab:table1}
\end{table*}

Table~\ref{tab:table1} summarizes the resulting magnetic-background domains. On the Maxwell-connected branch, $\mathcal{L}_F>0$ and $\mathcal C>0$ ensure a real, nondegenerate characteristic cone and a locally invertible Legendre map. The boundary $\mathcal C=0$ is excluded from these domains.

\section{Finite-amplitude nonlinear evolution} \label{sec:6}

The characteristic and linear analyses of the previous sections describe the local propagation properties of perturbations around a prescribed background, but they do not determine the evolution of a finite-amplitude waveform once different portions of the wave propagate with different characteristic velocities. In order to address this regime, we solve the full nonlinear field equations without expanding in the wave amplitude.

We shall restrict our attention to a one-dimensional propagation of the electromagnetic wave, along the $x$ direction. Since the electric field is transverse to the direction of propagation, we have $E\equiv E_y$, and the nonlinear Maxwell equations in Eqs. \eqref{eq:NLEDfieldeq1}--\eqref{eq:NLEDfieldeq4} reduce to
\begin{align}
\p_t D+\p_x H&=0\ , \label{eq:dDdt}\\
\p_t B+\p_xE&=0\ ,\label{eq:dBdt}
\end{align}
where $D=\mathcal{L}_F E$ and $H=\mathcal{L}_F B$.\footnote{The same reduced equations arise in a $\paren{3+1}$-dimensional $\mathcal{L}(F)$ theory for fields depending on $t$ and $x$, with $\vE=E_y\hat{\vy}$ and $\vB=B_z\hat{\vz}$. In the present $\paren{2+1}$-dimensional theory, $B=F_{12}$ is instead the complete magnetic field, rather than a selected component of a three-dimensional magnetic axial sector.} We study the finite-amplitude evolution of this system through exact simple-waves and an independent numerical solution.

For this analysis, it is convenient to introduce the following dimensionless variables using the nonlinear scale $b$:
\be
\mathcal{E}=\frac{E}{b}\ ,\quad \mathcal{B}=\frac{B}{b}\ , \quad d=\frac{D}{b}\ , \quad h=\frac{H}{b}\ , 
\en
and
\be
f=\frac{F}{b^2}=\frac{1}{2}\paren{\mathcal{E}^2-\mathcal{B}^2}\ , \quad \ell\paren{f}=\frac{\mathcal{L}\paren{F}}{b^2}\ .
\en
The three NLED models considered in this work have dimensionless Lagrangian densities given by
\begin{align}
\ell_\text{BI}\paren{f}&=1-\sqrt{1-2f}\ ,\label{eq:lBI} \\
\ell_{\log}\paren{f}&=-\ln\paren{1-f}\ , \label{eq:llog}\\
\ell_p\paren{f}&=f+\frac{f^2}{2}\ . \label{eq:lpoly}
\end{align}

We consider a homogeneous magnetic background $\mathcal{B}_0=\beta$, such that the magnetic perturbation can be written as $\delta\mathcal{B}=\mathcal{B}-\beta$. Hence we can write the system \eqref{eq:dDdt}--\eqref{eq:dBdt} in a conservation form as
\be \label{eq:consflux}
\p_t\vU+\p_x\vFc\paren{\vU}=0\ , 
\en
with
\be
\vU=\begin{pmatrix}
d\\
\delta\mathcal{B}
\end{pmatrix}\quad \text{and}\quad\vFc\paren{\vU}=\begin{pmatrix}
h-h_0\\
\mathcal{E}
\end{pmatrix}\ .
\en
In the latter we set the constant magnetic excitation of the background as $h_0=\beta\ell_f\paren{-\beta^2/2}$, in order for the homogeneous state $\vU=0$ to represent exact numerical equilibrium.

At each point of the evolution, the primitive electromagnetic variables, $\paren{\mathcal{E},\mathcal{B}}$ must be recovered from the variables contained in $\vU$. Thus we need to solve
\be
d=\ell_f\left[\frac{1}{2}\paren{\mathcal{E}^2-\mathcal{B}^2} \right]\mathcal{E}\quad\text{and}\quad\mathcal{B}=\beta+\delta\mathcal{B}\ .
\en
For the three models considered here, we select the real branch continuously connected to the Maxwell limit; the corresponding primitive-variable recovery relations are given in Appendix~\ref{ap:numerics}. Once $\paren{\mathcal{E},\mathcal{B}}$ are reconstructed, the magnetic excitation is obtained from $h=\ell_f\paren{f}\mathcal{B}$, and therefore the flux $\vFc\paren{\vU}$ is completely determined by the state $\vU$.

The local propagation properties of the nonlinear system follow directly from Eq. \eqref{eq:consflux}. Using the chain rule, we can write this equation as
\be 
\p_t\vU+\vA\paren{\vU}\p_x\vU=0\ ,
\en
where $\vA\paren{\vU}\equiv\p\vFc/\p\vU$ is the Jacobian matrix of the flux. The eigenvalues of $\vA$ determine the two local characteristic velocities of the system, namely
\be
\frac{dx}{dt}=\lambda_\pm\paren{\vU}\ .
\en
Thus, in contrast with Maxwell electrodynamics, the characteristic cone is in general state dependent: different portions of a finite-amplitude wave may propagate with different local characteristic velocities.

For a generic dimensionless NLED Lagrangian, $\ell\paren{f}$, the eigenvalues of the flux Jacobian are
\be \label{eq:lambdapm}
\lambda_\pm=\frac{\mathcal{E}\mathcal{B}\ell_{ff}\pm\sqrt{\ell_f\paren{\ell_f+2f\ell_{ff}}}}{\ell_f+\mathcal{E}^2\ell_{ff}}\ ,
\en
which for the homogeneous magnetic background, $\mathcal{E}=0$ and $f=-\beta^2/2$, reduces to
\be
\lambda_{\pm}=\pm\sqrt{\frac{\ell_f+2f\ell_{ff}}{\ell_f}}\ ,
\en
which is precisely the propagation velocity derived from the linear propagation analysis in Eq. \eqref{eq:disprelNLED}.

\subsection{Exact simple-wave evolution}

Before presenting the numerical results, it is useful to construct a class of exact solutions that allows the nonlinear propagation to be analyzed directly in terms of the characteristic structure of the system. We shall focus on right-moving simple-waves associated with the $\lambda_+$ characteristic family.

A simple-wave is a solution for which the two electromagnetic variables are not independent. In the present case we can parametrize the magnetic field in terms of the electric field as $\mathcal{B}=\mathcal{B}\paren{\mathcal{E}}$ such that $\mathcal{B}\paren{0}=\beta$. For a right-moving simple-wave, the electric field satisfies
\be \label{eq:124}
\p_t\mathcal{E}+\lambda_+\paren{\mathcal{E},\mathcal{B}}\p_x\mathcal{E}=0\ ,
\en
while Eq. \eqref{eq:dBdt} for the simple-wave implies
\be
\p_t\mathcal{B}=\frac{d\mathcal{B}}{d\mathcal{E}}\p_t\mathcal{E}\ .
\en
Hence, from the two latter equations, we have
\be \label{eq:dBdE}
\frac{d\mathcal{B}}{d\mathcal{E}}=\frac{1}{\lambda_+\paren{\mathcal{E},\mathcal{B}}}\ .
\en

Let the initial electric field profile be
\be
\mathcal{E}\paren{x,0}=\mathcal{E}_0\paren{\xi}\ ,
\en
where $\xi$ labels the initial position of each characteristic. For the simple-wave we have that the magnetic field is obtained from $\mathcal{B}\paren{x,0}=\mathcal{B}\left[\mathcal{E}_0\paren{\xi}\right]$. Next, the field equation takes the scalar form
\be
\p_t\mathcal{E}+\Lambda\paren{\mathcal{E}}\p_x\mathcal{E}=0\ ,
\en
where we have defined the characteristic velocity for the simple-wave as $\Lambda\paren{\mathcal{E}}\equiv\lambda_+\left[\mathcal{E},\mathcal{B}\paren{\mathcal{E}}\right]$. Notice that the electric field is constant along a characteristic satisfying
\[
\frac{dx}{dt}=\Lambda\paren{\mathcal{E}}\ .
\]
Consequently, the characteristic emerging from the initial point $\xi$ is a straight line,
\be \label{eq:char}
x\paren{\xi,t}=\xi+\Lambda\left[\mathcal{E}_0\paren{\xi}\right]t\ ,
\en
and, since $\mathcal{E}$ is constant along the characteristic, we have
\be \label{eq:charfield}
\mathcal{E}\left[x\paren{\xi,t},t\right]=\mathcal{E}_0\paren{\xi}\ .
\en
These two equations provide an exact implicit construction of the finite-amplitude simple-wave as long as the characteristic map remains single valued.

The nonlinear deformation of the waveform follows directly from the dependence of $\Lambda$ on the local field amplitude. Differentiating Eq. \eqref{eq:char} with respect to $\xi$ gives
\be
\frac{\p x}{\p \xi}=1+\Lambda'\left[\cE_0\paren{\xi}\right]\frac{\p\cE_0\paren{\xi}}{\p \xi}t\ ,
\en
where the prime on $\Lambda$ denotes derivative with respect to $\cE$.  Consequently, the spatial derivative of the electric field is
\be \label{eq:gradient-swave}
\p_x\cE=\frac{\paren{\p\cE_0/\p\xi}}{1+\Lambda'\left[\cE_0\paren{\xi}\right]\paren{\p\cE_0/\p\xi}t}\ .
\en
The behavior of the denominator in Eq. \eqref{eq:gradient-swave} can be used to determine the behavior of the wave as temporal evolution takes place. Whenever
\be
\Lambda'\left[\cE_0\paren{\xi}\right]\frac{\p\cE_0}{\p\xi}<0
\en
for some portion of the initial profile, neighboring characteristics converge and the magnitude of the field gradient grows with time. When the denominator vanishes a gradient catastrophe occurs, within a time given by
\be \label{eq:tshock}
t_\text{sh}=-\frac{1}{\min_\xi\left\{\Lambda'\left[\cE_0\paren{\xi}\right]\paren{\p\cE_0/\p\xi}\right\}}\ .
\en
If $\xi_*$ denotes the point where this minimum is attained, the position at which the catastrophe first develops is
\be
x_\text{sh}=\xi_*+\Lambda\left[\cE_0\paren{\xi_*}\right]t_\text{sh}\ .
\en

At $t=t_\text{sh}$ the field itself remains finite, whereas its spatial derivative formally diverges. Equivalently, $\p x/\p\xi$ vanishes and the characteristic in Eq. \eqref{eq:char} ceases to be invertible. Beyond this point the classical characteristic construction becomes multivalued and no longer represents a smooth solution of the original system. 

This construction provides a direct criterion for distinguishing exceptional from genuinely nonlinear propagation. If the characteristic velocity is constant along the simple-wave curve,
\be
\Lambda'\paren{\cE}=0\ ,
\en
and all characteristics belonging to the same family remain parallel, so no gradient catastrophe is generated by self-steepening. On the other hand, when $\Lambda'\ne0$, different portions of the waveform propagate at different velocities and characteristic compression may occur. The characteristic map predicts the smooth evolution and determines the onset of a gradient catastrophe through Eq.~\eqref{eq:tshock}.

The initial variation of the simple-wave velocity can be expressed directly in terms of the Lagrangian. Let $f_0=-\beta^2/2$ and $\mathcal{C}_0=\ell_f(f_0)-\beta^2\ell_{ff}(f_0)>0$. Evaluating Eq. \eqref{eq:dBdE} at the magnetic background and differentiating Eq. \eqref{eq:lambdapm} along the simple-wave curve gives
\be \label{eq:generalLambdaPrime}
\Lambda'(0)=
\frac{\beta^3}{2\ell_f(f_0)\mathcal C_0}
\left[\ell_f(f_0)\ell_{fff}(f_0)-3\ell_{ff}^2(f_0)\right].
\en
For the weak-field expansion in Eq.~\eqref{eq:Lgen}, this reduces to
\be
\Lambda'_i(0)=3\left(c_i-\frac12\right)\beta^3+\mathcal O\paren{\beta^5}.
\en
Thus, the first coefficient at which the three Lagrangians differ also controls their leading difference in simple-wave steepening on a weak magnetic background. The Born-Infeld coefficient gives a vanishing slope, whereas the logarithmic and polynomial coefficients give negative slopes for $\beta>0$, consistent with the exact model-specific expressions below.

To check the exact solutions during their smooth evolution, we also solve Eq.~\eqref{eq:consflux} with a finite-volume scheme. The scheme uses piecewise-linear MUSCL reconstruction with an MC limiter, a local Rusanov flux, and SSP-RK3 time integration with $C_{\mathrm{CFL}}=0.4$. We impose periodic boundary conditions on a domain large enough that the wave packet does not encounter its periodic image during the reported evolution. The full discretization and tabulated peak-gradient comparisons are given in Appendix~\ref{ap:numerics}.

\subsection{Born-Infeld exceptional propagation}

Let us apply the simple-wave construction to Born-Infeld electrodynamics. The Born-Infeld Lagrangian in dimensionless variables is given in Eq. \eqref{eq:lBI}, and we have the derivatives
\be
\ell_f=\frac{1}{\sqrt{1-2f}}\ , \quad \ell_{ff}=\frac{1}{\paren{1-2f}^{3/2}}\ .
\en
The right-moving characteristic velocity in Eq. \eqref{eq:lambdapm} reduces to
\be
\lambda_+^\text{BI}\paren{\cE.\cB}=\frac{\cE\cB+\sqrt{1-\cE^2+\cB^2}}{1+\cB^2}\ ,
\en
which for the homogeneous magnetic background, $\cE=0$, $\cB=\beta$, results in
\be
v_{0,\text{BI}}\equiv\lambda_+^{\text{BI}}\paren{0,\beta}=\frac{1}{\sqrt{1+\beta^2}}\ .
\en
Thus, we can readily solve Eq. \eqref{eq:dBdE}, obtaining
\be
\cB\paren{\cE}=\beta+\sqrt{1+\beta^2}\cE\ .
\en
In particular, the latter equations imply that $\lambda_+^{\text{BI}}=\Lambda_\text{BI}\paren{\cE}=1/\sqrt{1+\beta^2}$ everywhere, hence
\be
\Lambda'_\text{BI}\paren{\cE}=0\ .
\en
Thus, the characteristics belonging to this simple-wave family propagate with the same velocity, independently of the local wave amplitude. Henceforth, the characteristics obey
\be
x\paren{\xi,t}=\xi+v_{0,\text{BI}}t,
\en
which gives the exact electric and magnetic fields:
\begin{align}
\cE\paren{x,t}&=\cE_0\paren{x-v_{0,\text{BI}}t}\ , \\
\cB\paren{x,t}&=\beta+\sqrt{1+\beta^2}\cE_0\paren{x-v_{0,\text{BI}}t}\ .
\end{align}

Thus, an arbitrary smooth profile belonging to this right-moving simple-wave is transported rigidly without nonlinear self-steepening. This can also be directly proven by calculating the shock time in Eq. \eqref{eq:tshock} for this theory. In particular, $\Lambda'_{\text{BI}}\paren{\cE}=0$ implies that no gradient catastrophe occurs at finite time, so  $t_\text{sh}=\infty$ for Born-Infeld electrodynamics. These results illustrate the exceptional propagation of Born-Infeld theory \cite{Kadlecova2024}. The full $\paren{3+1}$-dimensional theory also exhibits no birefringence \cite{Boillat1970}.

\subsection{Gradient catastrophe in logarithmic and polynomial electrodynamics}

In contrast to the exceptional propagation found for Born-Infeld electrodynamics, in both logarithmic and polynomial models studied here, the characteristic velocity varies along the right-moving simple-wave. Thus, for these models $\Lambda'\paren{\cE}\ne0$, and different portions of the finite-amplitude pulse propagate with different velocities. 

In order to demonstrate the existence of such a phenomenon, let us consider a smooth Gaussian electric field profile of the form
\be
\cE_0\paren{\xi}=A\exp\left[-\frac{\paren{\xi-\xi_c}^2}{2\sigma^2} \right]\ ,
\en
and the initial magnetic field is determined from the exact relation $\cB_0\paren{\xi}=\cB\left[\cE_0\paren{\xi} \right]$. We now apply these to both logarithmic and polynomial models of electrodynamics.

\subsubsection{Logarithmic electrodynamics}
For logarithmic electrodynamics, the dimensionless Lagrangian density is given in Eq. \eqref{eq:llog} and its derivatives are
\be
\ell_{f}=\frac{1}{1-f}\ , \quad \ell_{ff}=\frac{1}{\paren{1-f}^2}\ .
\en
Substitution into the general characteristic velocity in Eq. \eqref{eq:lambdapm} gives
\be
\lambda_{+}^{\log}=\frac{\cE\cB+\sqrt{1-f^2}}{1+\frac{1}{2}\paren{\cE^2+\cB^2}}\ .
\en
The corresponding simple-wave is obtained by solving 
\be
\frac{d\cB}{d\cE}=\frac{1}{\lambda^{\log}_+\paren{\cE,\cB}}\ , \quad \cB\paren{0}=\beta\ ,
\en
Unlike the Born-Infeld case, the resulting velocity is not constant along the simple-wave curve. This departure from linear degeneracy can already be seen in the weak-amplitude limit. Expanding the characteristic velocity along the simple-wave around the magnetic background gives
\be\label{eq:lambdaprimeLog}
\Lambda'_{\log}\paren{0}=-\frac{2\beta^3}{4-\beta^4}\ ,
\en
for which $\Lambda'_{\log}<0$ for $\beta<\sqrt{2}$, i.e., $\Lambda'_{\log}<0$ within the hyperbolicity domain (see Table \ref{tab:table1}). The negative sign in $\Lambda'_{\log}$ implies that, for a positive pulse, larger values of the electric field propagate more slowly. Consequently, the rising side of a right-moving pulse, for which $\paren{\p \cE_0}/\paren{\p\xi}>0$, satisfies
\be
\Lambda'_{\log}\frac{\p\cE_0}{\p\xi}<0\ , 
\en
and undergoes characteristic compression.

The complete finite-amplitude shock time is obtained by numerically evaluating Eq. \eqref{eq:tshock} for the theory under inspection. 

The characteristic and finite-volume gradient comparisons are shown in Fig.~\ref{fig:fig2} and tabulated in Appendix~\ref{ap:numerics}.

It is important to emphasize that the steepening observed here does not result from a loss of hyperbolicity. The dimensionless hyperbolicity combination $\mathcal{C}\equiv\ell_f+2f\ell_{ff}$ is, for the logarithmic model, 
\be
\mathcal{C}_{\log}=\frac{1+f}{\paren{1-f}^2}>0\quad\text{for }f>-1\ .
\en
For the evolution above, even at $t=0.9t_{\text{sh}}$ we find
\be
\min\paren{1+f}\simeq0.460>0\ ,
\en
so that the solution remains strictly hyperbolic while its spatial gradient grows rapidly.

\subsubsection{Polynomial electrodynamics}

For the polynomial model in Eq. \eqref{eq:lpoly} we have the derivatives
\be
\ell_f=1+f\ , \quad \ell_{ff}=1\ .
\en
In this case, the right-moving characteristic velocity is
\be
\lambda_+^p=\frac{\cE\cB+\sqrt{\paren{1+f}\paren{1+3f}}}{1+f+\cE^2}\ .
\en
Near the homogeneous background, the variation of the characteristic velocity is 
\be \label{eq:lambdaprimePoly}
\Lambda'_p\paren{0}=-\frac{6\beta^3}{\paren{2-\beta^2}\paren{2-3\beta^2}}\ ,
\en
which is negative for $\beta^2<2/3$, which in turn gives the hyperbolicity domain for the theory. The finite-volume solution again follows the characteristic prediction during smooth evolution (Fig.~\ref{fig:fig2}); numerical values of the peak gradient are given in Appendix~\ref{ap:numerics}.

For this theory, the hyperbolicity combination results in $\mathcal{C}_p=1+3f$, thus hyperbolicity requires $f>-1/3$. For instance, at $t=0.9 t_\text{sh}^p$, the numerical evolution gives approximately $\min\paren{1+3f}\simeq0.209>0$. Thus, also in this case, the rapid growth of the field gradient occurs while the system remains strictly hyperbolic.

\begin{figure*}[t]
\centering
\includegraphics[width=0.9\textwidth]{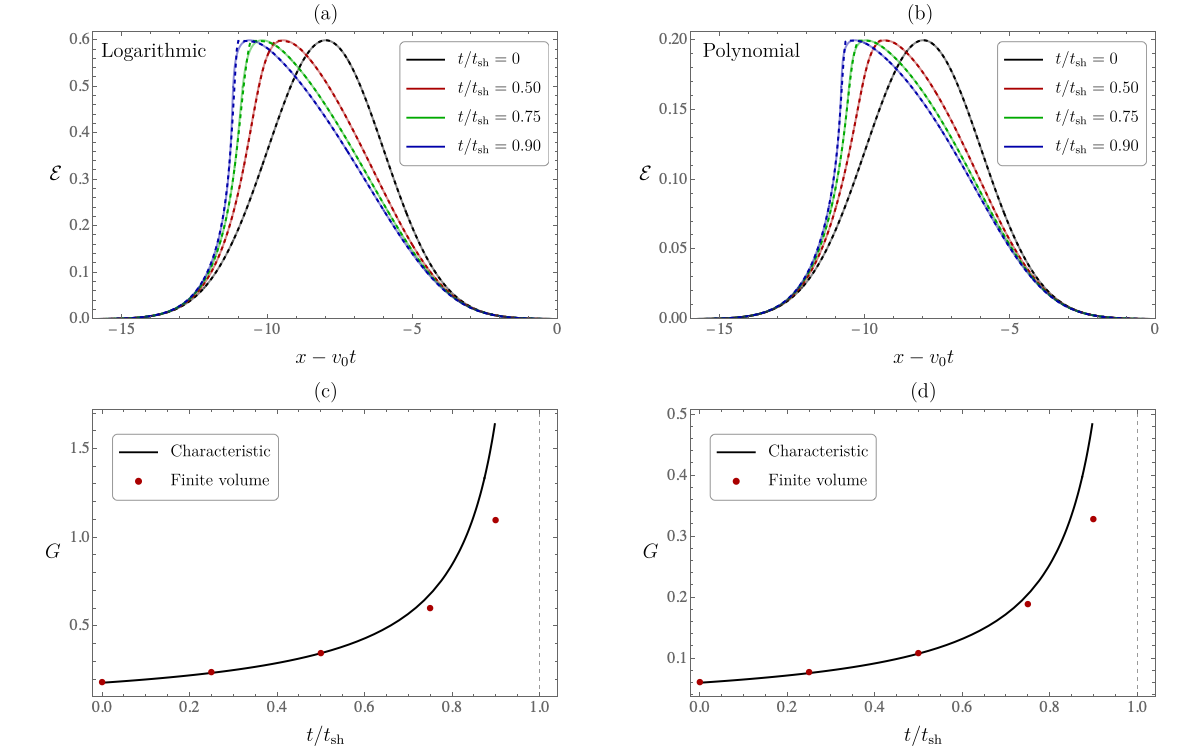}
\caption{Right-moving simple-wave in logarithmic $\paren{\text{a,c}}$ and polynomial $\paren{\text{b,d}}$ electrodynamics, with $\beta=0.5$, $\sigma=2$, and $A=0.6$ and $0.2$ respectively. Upper panels show $\cE$ as function of $x-v_0t$; solid curves represent characteristic solutions, and dashed curves represent finite-volume solutions. Lower panels show the gradient $G=\max_x\vert\p_x\cE\vert$ as a function of $t/t_{\text{sh}}$; curves represent characteristic predictions while points represent finite-volume results, and vertical dashed lines mark $t_{\text{sh}}$.}
\label{fig:fig2}
\end{figure*}

Fig. \ref{fig:fig2} presents two independent validation cases with different pulse amplitudes. In both, the finite-volume profiles follow the characteristic solutions while smooth, whereas the peak gradients fall below the characteristic prediction near $t_{\text{sh}}$ because the numerical scheme regularizes the steepening front.

The growth of the spatial gradient and the loss of hyperbolicity correspond to distinct phenomena. In both logarithmic and polynomial electrodynamics, the simple-wave evolution develops increasingly strong characteristic compression while the corresponding hyperbolicity condition remains strictly satisfied. Thus, the gradient catastrophe is generated by the amplitude dependence of the characteristic velocity, which causes neighboring characteristics to converge, rather than by an approach to the boundary $\mathcal{C}=0$. In particular, the electromagnetic state carried by each characteristic remains finite and within the hyperbolic sector up to the first characteristic crossing, even though $\left|\p_x\cE\right|$ diverges formally in the exact solution.

Taken together with the Born-Infeld result of the previous subsection, these calculations reveal qualitatively different finite-amplitude propagation within theories that share the same weak-field expansion. Born-Infeld electrodynamics is linearly degenerate along the simple-wave branch considered here and therefore exhibits rigid propagation, whereas the logarithmic and polynomial models possess amplitude-dependent characteristic velocities and develop finite-time gradient catastrophes. We next place the three theories on equal footing by evolving simple-waves with the same background, amplitude, and width, allowing their finite-amplitude dynamics to be compared directly.

\subsection{Matched finite-amplitude comparison}

The results obtained above show that the three weak-field-matched theories belong to qualitatively different dynamical classes. Born-Infeld electrodynamics possesses exceptional propagation along the simple-wave, whereas logarithmic and polynomial electrodynamics develop finite-time gradient catastrophes. In order to compare these differences directly, we now prescribe the same electric field pulse in all three theories and evolve it over the same physical time.

We shall consider the common parameters $\beta=0.5$, $A=0.2$, and $\sigma=2$, for an initial electric field of the form
\be
\cE_0\paren{x}=A\exp\left[-\frac{\paren{x-x_c}^2}{2\sigma^2}\right]\ , 
\en
where $x_c$ is the initial pulse center. The electric field profile is therefore identical in the three theories. However, the magnetic field for each theory is determined separately from the exact simple-wave relation of each theory, which has the generic form
\[
\cB_0^{(i)}\paren{x}=\cB^{(i)}\left[ \cE_0\paren{x}\right]\ ,
\]
where the index $i$ stands for the theory under consideration. This prescription ensures that the initial data belong to the same characteristic family in each nonlinear theory, rather than imposing an identical magnetic perturbation that would in general excite both characteristic modes.

For these parameters, the characteristic analysis gives noticeably different gradient catastrophe times, namely
\be
t_{\text{sh}}^{\text{BI}}=\infty,\quad t_{\text{sh}}^{\log}\simeq222.85,\quad t_{\text{sh}}^{\text{p}}\simeq17.89\ .
\en 
We therefore choose a common comparison time preceding the earliest gradient catastrophe,
\be
t_{\text{cmp}}=0.75 t_{\text{sh}}^p\simeq13.42\ .
\en
For this choice, the polynomial pulse has already undergone a substantial fraction of its evolution before attaining the gradient catastrophe, while for the logarithmic model $t_{\text{cmp}}/t_{\text{sh}}^{\log}\simeq0.060$.

\begin{figure*}[t]
\centering
\includegraphics[width=0.9\textwidth]{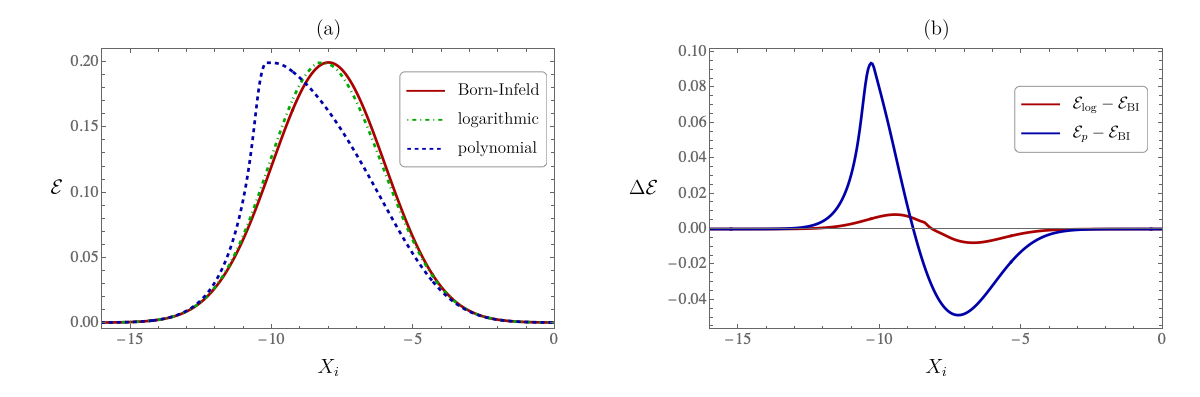}
\caption{Evolution of right-moving simple-waves in Born-Infeld, logarithmic, and polynomial electrodynamics from the same Gaussian electric pulse ($\beta=0.5$, $A=0.2$, $\sigma=2$), shown at $t_{\text{cmp}}=0.75t_{\text{sh}}^p$. Panel (a) shows electric profiles in the comoving coordinates $X_i=x-v_{0,i}t$; panel (b) shows $\cE_{\log}-\cE_{\text{BI}}$ and $\cE_p-\cE_{\text{BI}}$ at equal comoving coordinates. The initial magnetic profiles satisfy each model's simple-wave relation.}
\label{fig:fig3}
\end{figure*}

A direct comparison in the original spatial coordinate would also contain the model dependence of the background propagation velocity. For $\beta=0.5$ the velocities are
\be
v_{0,\text{BI}}\simeq0.89443\ ,\quad v_{0,\log}\simeq0.88192\ ,\quad v_{0,p}\simeq0.84515\ .
\en
To isolate the finite-amplitude deformation from this trivial difference in the background motion, we introduce the corresponding background comoving coordinate for each theory
\be
X_i=x-v_{0,i}t\ ,
\en
where $i$ stands for the theory under consideration. Fig. \ref{fig:fig3} shows the resulting profiles at $t=t_{\text{cmp}}$. The Born-Infeld pulse retains its initial Gaussian shape, as expected from the exact result $\Lambda'_{\text{BI}}=0$. The logarithmic solution remains close to the Born-Infeld profile, although a small displacement and deformation are already visible. By contrast, the polynomial pulse displays a pronounced shift toward smaller $X_p$ together with strong steepening of its rising flank and broadening of the opposite side. These features follow directly from the stronger dependence of $\Lambda_p$ on the local field amplitude: the high-field portion of the pulse propagates more slowly than the low-amplitude tails, producing rapid accumulation of characteristics.

In panel (b) of Fig. \ref{fig:fig3} we display the departures of both logarithmic and polynomial electrodynamics with respect to the Born-Infeld case. Because the Born-Infeld pulse propagates rigidly in its comoving coordinate, its profile coincides with the initial Gaussian pulse and therefore provides a natural reference for the nonlinear deformation of the other two theories. This plot reflects both the backward displacement of the pulse relative to the initial wave profile and the redistribution of its spatial profile produced by nonlinear steepening.

The comparison in Fig.~\ref{fig:fig3} demonstrates that substantial differences among the three theories are already present for the same initial electric pulse on a common magnetic background. It is therefore natural to ask how these differences evolve as the pulse amplitude is varied. We now address this question systematically by studying the amplitude dependence of the gradient-catastrophe time and of the hyperbolicity conditions along the corresponding simple-wave branches.

\begin{figure*}[t]
\centering
\includegraphics[width=0.9\textwidth]{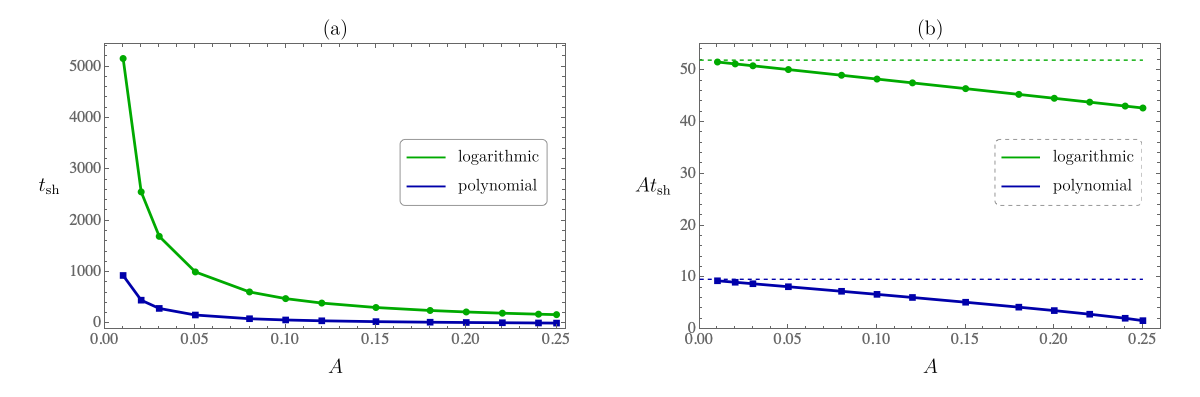}
\caption{Characteristic prediction of the gradient-catastrophe time of logarithmic and polynomial right-moving simple-waves at $\beta=0.5$ and $\sigma=2$. Panels (a) and (b) show $t_{\text{sh}}$ and $A t_{\text{sh}}$ as a function of the amplitude $A$, respectively. Horizontal dashed lines mark the weak-amplitude limits $A t_{\text{sh}}\to\sigma\sqrt{e}/\vert\Lambda'\paren{0}\vert$. }
\label{fig:fig4}
\end{figure*}

Let us first examine how the gradient-catastrophe time depends on the amplitude of the initial pulse while keeping the magnetic background and the pulse width fixed at $\beta=0.5$ and $\sigma=2$, respectively. Panel (a) of Fig. \ref{fig:fig4} shows $t_{\text{sh}}$ as a function of $A$ for logarithmic and polynomial electrodynamics. In both theories, the catastrophe time increases rapidly as the amplitude is reduced and diverges in the limit $A\rightarrow0$. This behavior is expected, since in the weak-amplitude limit the system approaches linear propagation and the amplitude-dependent variation of the characteristic velocity becomes increasingly smaller. However, for other values of the amplitude, the polynomial model develops a gradient catastrophe considerably earlier than the logarithmic theory.

To determine the weak-amplitude limit of Eq. \eqref{eq:tshock}, we expand the characteristic velocity along the simple-wave branch around the homogeneous background, hence
\be
\Lambda\paren{\cE}=v_0+\Lambda'\paren{0}\cE+\mathcal{O}\paren{\cE^2}\ .
\en
For the Gaussian initial profile, the maximum magnitude of the relevant initial gradient is
\be
\max_\xi\left|\frac{\p\cE_0}{\p\xi} \right|=\frac{A}{\sigma\sqrt{e}}\ .
\en
Consequently, the weak-amplitude gradient-catastrophe time behaves as 
\be
t_\text{sh}\simeq\frac{\sigma\sqrt{e}}{\vert\Lambda'\paren{0}\vert A}\ ,
\en
thus, as $A\to0$, we have $A t_\text{sh}\approx\sigma\sqrt{e}/\vert\Lambda'\paren{0}\vert$. For $\beta=0.5$ and $\sigma=2$, Eqs. \eqref{eq:lambdaprimeLog} and \eqref{eq:lambdaprimePoly} give
\be \label{eq:Atsh-exact}
A t_\text{sh}^{\log}\approx 51.9\ , \quad At_\text{sh}^p\approx9.62\ .
\en

Panel (b) of Fig. \ref{fig:fig4} shows $A t_\text{sh}$ as a function of $A$ for both logarithmic and polynomial electrodynamics. The asymptotic values in Eq. \eqref{eq:Atsh-exact} are displayed as dashed horizontal lines. The progressive departure of $A t_\text{sh}$ from these constant values as the amplitude grows reflects the finite-amplitude corrections beyond the leading order expansion of $\Lambda\paren{\cE}$ around the homogeneous background.

\begin{figure}[t]
\centering
\includegraphics[width=0.45\textwidth]{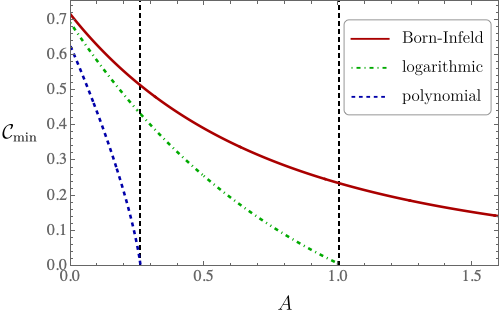}
\caption{Minimum hyperbolicity margin $\mathcal{C}_{\min}$ along the exact right-moving simple-wave branch as a function of the electric-field amplitude $A$, for $\beta=0.5$. Vertical dashed lines mark the critical amplitudes of the polynomial and logarithmic models, where $\mathcal{C}_{\min}=0$.}
\label{fig:fig5}
\end{figure}

As discussed previously, the development of a gradient catastrophe does not require loss of hyperbolicity, and it becomes important to assess this last feature for the finite-amplitude simple-wave propagation. In order to address the loss of hyperbolicity, we evaluate the minimum value of the combination $\mathcal{C}\equiv\ell_f+2f\ell_{ff}$ for which hyperbolicity holds when $\mathcal{C}>0$. We denote this minimum value as $\mathcal{C}_{\min}$. Fig. \ref{fig:fig5} shows this quantity as a function of $A$ for the three theories at $\beta=0.5$. As $A\to0$, $\mathcal{C}_{\min}$ reduces to its value on the homogeneous magnetic background, and as $A$ increases, the three theories display markedly different behavior. For the polynomial model, the hyperbolicity margin decreases rapidly and reaches zero at
\be
A_\text{crit}^p\simeq0.262\ ,
\en
while logarithmic electrodynamics remains hyperbolic until $A_\text{crit}^{\log}\simeq1.0$. By contrast, Born-Infeld electrodynamics remains hyperbolic for every finite positive amplitude.

It is worth emphasizing that $A_\text{crit}$ does not represent a shock-formation threshold. For amplitudes below $A_\text{crit}$, logarithmic and polynomial waves may still develop a gradient catastrophe at a finite time while remaining strictly hyperbolic, as discussed earlier. The critical amplitude $A_\text{crit}$ determines the largest amplitude for which the complete initial simple-wave profile lies within the hyperbolic sector of the corresponding theory. 

\begin{figure}[t]
\centering
\includegraphics[width=0.48\textwidth]{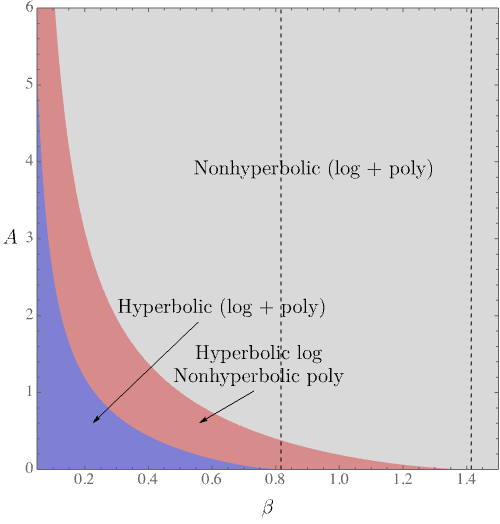}
\caption{Hyperbolicity domains of logarithmic and polynomial right-moving simple-waves in the $\paren{\beta,A}$ plane. The critical curves $A_{\text{crit}}^p\paren{\beta}$ and $A_{\text{crit}}^{\log}\paren{\beta}$ satisfy $\mathcal{C}_{\min}=0$. Vertical dashed lines mark the background limits $\beta_{\max}^p=\sqrt{2/3}$ and $\beta_{\max}^{\log}=\sqrt{2}$.}
\label{fig:fig6}
\end{figure}

The critical amplitude also depends strongly on the magnetic background. To display this dependence globally, Fig. \ref{fig:fig6} shows the curves $A_\text{crit}^p\paren{\beta}$ and $A_\text{crit}^{\log}\paren{\beta}$ in the $\paren{\beta,A}$ plane. These curves divide the parameter space into regions with different hyperbolicity properties. Below the polynomial critical curve (purple region), both logarithmic and polynomial simple-waves are hyperbolic. Between the polynomial and logarithmic boundaries (red region), the logarithmic theory remains hyperbolic while the polynomial model does not. Above the logarithmic boundary, neither of the two theories is hyperbolic for the simple-waves considered here. The corresponding homogeneous-background hyperbolicity limits are  $\beta=\sqrt{2}$ and $\beta=\sqrt{2/3}$, respectively. Conversely, as the homogeneous magnetic field decreases, higher amplitude waves are allowed before losing hyperbolicity. Born-Infeld electrodynamics is not shown because the right-moving simple-wave branch considered here has no finite-amplitude hyperbolicity boundary.

\section{Concluding remarks} \label{sec:concl}

In this work, we have investigated wave propagation in nonlinear electrodynamics in $\paren{2+1}$ dimensions, focusing on three theories whose Lagrangian densities share the same weak-field expansion through the first nonlinear correction: Born-Infeld, logarithmic, and polynomial electrodynamics. This construction provides a controlled setting in which models that are indistinguishable at the first nonlinear order can be compared beyond the perturbative regime. Our analysis combined the characteristic structure of the field equations, physical admissibility conditions, exact simple-wave solutions, and finite-volume evolutions in order to determine when differences between the theories become dynamically relevant.

When considering propagation over a homogeneous background, the three theories already exhibit different characteristic velocities once terms involving the derivatives of the Lagrangian density become important. We have also considered other physical viability criteria for these theories, such as causal propagation, hyperbolicity, a one-to-one Legendre mapping and energy conditions. We find that the combination $\mathcal{L}_F+2F\mathcal{L}_{FF}$, together with $\mathcal{L}_F>0$, governs both magnetic-background hyperbolicity and local invertibility of the Legendre map. This connection has also appeared in the context of regular black hole solutions \cite{Bronnikov2001}. 

Beyond weak-amplitude perturbations, simple-wave propagation reveals further differences among the theories. In this regime, the characteristic velocity is no longer determined only by the background field, but varies along the nonlinear state curve explored by the wave. In this sense, different portions of the same waveform effectively experience different local propagation cones, and the variation of these cones across the pulse determines whether the profile propagates rigidly, spreads, or undergoes characteristic compression.

In this respect, Born-Infeld electrodynamics is an exceptional theory: although its propagation velocity differs from the Maxwell value in the presence of a magnetic background, the velocity remains constant along the exact simple-wave. Thus, the nonlinear constitutive response changes the propagation speed, but all portions of the wave travel at that same speed, and no self-steepening or gradient catastrophe occurs at finite time. The exceptional character of Born-Infeld theory was established by Boillat using the characteristic propagation of weak-discontinuities \cite{Boillat1970}; the exact simple-wave construction presented here provides a complementary finite-amplitude, nonperturbative realization of this property within the propagation sector considered.

By contrast, logarithmic and polynomial electrodynamics are genuinely nonlinear along their simple-wave branches. Their characteristic velocities depend on the local field amplitude, so different portions of the pulse propagate at different speeds and produce characteristic compression. The resulting gradient catastrophe is associated with the crossing of neighboring characteristics rather than with a divergence of the field itself. This effect is considerably stronger in the polynomial model, which develops a much shorter catastrophe time than the logarithmic theory.

Gradient catastrophe and loss of hyperbolicity are nevertheless distinct phenomena. The former is controlled by the variation of the characteristic velocity along the nonlinear state curve, whereas the latter occurs when the hyperbolicity condition $\mathcal{C}=\ell_f+2f\ell_{ff}>0$ is violated. A wave may therefore develop a divergent spatial gradient while remaining entirely within the hyperbolic sector. The polynomial theory combines faster steepening with a smaller admissible finite-amplitude domain, while the logarithmic model remains hyperbolic over a wider region of parameter space.

Overall, our results show that agreement in the weak-field expansion does not determine the finite-amplitude dynamics of a nonlinear electromagnetic theory. Once the full constitutive structure is probed, weak-field-matched models can display qualitatively different behavior, ranging from exceptional rigid propagation to finite-time self-steepening. In the weak-background regime, this distinction can be traced directly to the first higher-order coefficient at which the matched Lagrangians differ, which controls the leading amplitude dependence of the simple-wave characteristic velocity. Finite-amplitude wave propagation therefore provides a complementary way of distinguishing NLED theories beyond their perturbative expansions. A natural extension is to consider more general Lagrangians and configurations involving both characteristic families, such as collisions of counter-propagating waves, which we intend to address in future works.

\appendix

\section{The $\zeta=0$ sector} \label{ap:A}

Equation \eqref{eq:37} is a necessary consequence of the discontinuity equations, obtained after contracting them. Its factor $\zeta=0$ therefore cannot be interpreted as an additional characteristic relation. Let us assume that $\mathcal{L}_F\ne0$, as on the Maxwell-connected branch considered in the main text.

In three spacetime dimensions, an antisymmetric tensor is dual to a vector. We define the duals of the background fields and the discontinuity amplitude by
\be
\widetilde F^{a}\equiv\frac{1}{2}\epsilon^{abc}F_{bc},
\qquad
h^{a}\equiv\frac{1}{2}\epsilon^{abc}f_{bc},
\qquad
\epsilon^{012}=+1.
\en
This duality is invertible, so $f^{ab}=\epsilon^{abc}h_c$. It also gives $\zeta=F^{ab}f_{ab}=2\widetilde F_a h^a$. Contracting the discontinuous Bianchi identity, Eq.~\eqref{eq:dBianchi}, with $\epsilon^{abc}$ yields
\be
k_a h^a=0.
\en

Now set $\zeta=0$ in the discontinuous field equation, Eq.~\eqref{eq:dField}. Since $\mathcal L_F\ne0$, it reduces to
\be
k_a f^{ab}=k_a\epsilon^{abc}h_c=0.
\en
The last equality is equivalent to $k_{[a}h_{b]}=0$: the vectors $k^a$ and $h^a$ are linearly dependent. Thus $h^a=\alpha k^a$. For a nontrivial discontinuity, $f_{ab}\ne0$ and hence $h^a\ne0$, so $\alpha\ne0$. Substituting this relation into $k_a h^a=0$ and $\zeta=2\widetilde F_a h^a=0$ gives two necessary conditions:
\be
k_a k^a=0,
\qquad
\widetilde F_a k^a=0.
\en
A possible $\zeta=0$ discontinuity must therefore have a Minkowski-null wave covector orthogonal to the dual background field.

For the homogeneous magnetic background used in Sec.~\ref{sec:4}, $F_{12}=B_0$ and the electric field vanishes. Consequently, $\widetilde F^a=(B_0,0,0)$. When $B_0\ne0$, the orthogonality condition sets $k^0=0$. The null condition then reads $0=k_a k^a=-k_x^2-k_y^2$, forcing $k_x=k_y=0$. Hence $k_a=0$, which cannot be the normal to a characteristic hypersurface. There is therefore no nontrivial $\zeta=0$ mode on a nonzero homogeneous magnetic background.

At $B_0=0$, $\zeta$ vanishes identically, while the jump equations give $k_a k^a=0$. This is the ordinary Maxwell cone and coincides with Eq.~\eqref{eq:geff} in the zero-field limit. Thus the $\zeta=0$ factor supplies no second propagation cone for the backgrounds considered here.

\section{Details of the admissibility analysis} \label{ap:admissibility}

\subsection{Inverse constitutive relations}

The invariant map $P=F\mathcal{L}_F^2$ introduced in Eq.~\eqref{eq:invP} is locally invertible wherever $dP/dF=\mathcal{L}_F\mathcal C$ is nonzero. On the connected magnetic-background domains considered here, $\mathcal{L}_F>0$ and $\mathcal C>0$ make this map strictly increasing. Let us now construct the inverse formulation by defining the Hamiltonian-like density
\be
\mathcal{H}\paren{P}=2F\mathcal{L}_F-\mathcal{L}\paren{F}\ .
\en
Differentiating with respect to $F$ and using Eq.~\eqref{eq:dPdF} gives
\be
\frac{d\mathcal{H}}{dF}=\mathcal C,\qquad
\mathcal{H}_P=\frac{d\mathcal{H}/dF}{dP/dF}=\frac{1}{\mathcal{L}_F}\ .
\en
Therefore, the inverse invariant and constitutive relations are
\be
F=P\mathcal{H}_P^2,\qquad F^{ab}=\mathcal{H}_P E^{ab}\ .
\en
The $F$ and $P$ formulations are therefore locally equivalent on a monotonic branch. This inverse construction fails at $\mathcal C=0$ or $\mathcal{L}_F=0$, where $dP/dF=0$. Related branching of the constitutive map has been discussed in Refs.~\cite{Bronnikov2001,Bronnikov2018}.

\subsection{Energy conditions on magnetic backgrounds}

Energy conditions are often used to assess NLED matter sources in gravitational applications \cite{Penafiel2017,Russo2024}. For the homogeneous magnetic background, the energy-momentum tensor has the form $T^a{}_{b}=\operatorname{diag}\paren{\rho,-p,-p}$, with $\rho=-\mathcal{L}$ and $p=\mathcal{L}-2F\mathcal{L}_F$. We now derive the conditions quoted in Sec.~\ref{sec:enCond}.

The null energy condition (NEC) requires $\rho+p\ge0$, which becomes
\be \label{eq:NECNLED}
-2F\mathcal L_F\ge0\ .
\en
The weak energy condition (WEC) further requires $\rho\ge0$, or $\mathcal{L}\le0$. The dominant energy condition (DEC) requires a nonnegative energy density and a causal energy flux. For the isotropic tensor above, this becomes $\rho\ge|p|$, or equivalently the pair $\rho+p\ge0$ and $\rho-p\ge0$. In terms of the Lagrangian, the additional inequality is
\be
\rho-p=2\paren{F\mathcal{L}_F-\mathcal{L}}\ge0\ .
\en
In $\paren{2+1}$ dimensions the strong energy condition (SEC) is
\be \label{eq:SEC}
\paren{T_{ab}-T\eta_{ab}}u^au^b\ge0\ ,
\en
for every timelike vector $u^a$, where $T=T^a{}_{a}=\rho-2p$ is the trace. For the isotropic magnetic stress tensor, the SEC reduces to $\rho+p\ge0$ and $p\ge0$. Since $F<0$ and $\mathcal{L}_F>0$ on the nonzero magnetic backgrounds considered here, the NEC is automatically satisfied. We therefore check $\rho$, $p$, and $\rho-p$ for each model.

Writing $x=\beta^2=B_0^2/b^2$, Born-Infeld electrodynamics gives
\begin{align}
\rho_\text{BI}&=b^2\paren{\sqrt{1+x}-1}\ ,\\
p_\text{BI}&=b^2\paren{1-\frac{1}{\sqrt{1+x}}}\ ,\\
\rho_\text{BI}+p_\text{BI}&=\frac{b^2x}{\sqrt{1+x}}\ ,\\
\rho_\text{BI}-p_\text{BI}&=\frac{b^2\paren{\sqrt{1+x}-1}^2}{\sqrt{1+x}}\ .
\end{align}
Each quantity is nonnegative for $x\ge0$, so all four energy conditions hold throughout the magnetic-background domain of this model.

For logarithmic electrodynamics, let $q=1+x/2$. The corresponding expressions are
\begin{align}
\rho_\text{log}&=b^2\ln q\ ,\\
p_\text{log}&=b^2\paren{\frac{x}{q}-\ln q}\ ,\\
\rho_\text{log}+p_\text{log}&=\frac{b^2x}{q}\ ,\\
\rho_\text{log}-p_\text{log}&=b^2\paren{2\ln q-\frac{x}{q}}\ .
\end{align}
The last expression vanishes at $x=0$ and has derivative $b^2x/(2q^2)\ge0$. Also $p_\text{log}\paren{0}=0$ and $dp_\text{log}/dx=b^2\paren{1-x/2}/\paren{2q^2}\ge0$ for $0\le x\le2$. The pressure reaches its maximum at $x=2$, the boundary of the hyperbolic domain. Thus the NEC, WEC, DEC, and SEC all hold for $0\le x<2$.

For the polynomial model we find
\begin{align}
\rho_p&=\frac{b^2x\paren{4-x}}{8}\ ,\\
p_p&=\frac{b^2x\paren{4-3x}}{8}\ ,\\
\rho_p+p_p&=b^2x\paren{1-\frac{x}{2}}\ ,\\
\rho_p-p_p&=\frac{b^2x^2}{4}\ .
\end{align}
In particular, $\rho_p\ge0$ for $x\le4$ and $p_p\ge0$ for $x\le4/3$. All four energy conditions therefore hold on the smaller hyperbolic domain $0\le x<2/3$. These results establish the energy-condition entries of Table~\ref{tab:table1}.

\section{Finite-volume scheme and gradient comparisons} \label{ap:numerics}

\subsection{Discretization}

In order to evolve Eq.~\eqref{eq:consflux}, we employ a finite-volume shock-capturing scheme based on the Monotonic Upstream-centered Schemes for Conservation Laws (MUSCL) approach \cite{vanLeer1979,LeVeque2002}. The computational domain is divided into cells of uniform width $\Delta x$, and $\vU_i$ denotes the cell-averaged conservative state in the $i$th cell. Rather than assuming the solution to be constant inside each cell, the MUSCL method reconstructs a piecewise-linear profile from the neighboring cell averages, thereby achieving second-order spatial accuracy in smooth regions.

To construct the local slope, we first introduce the backward and forward differences
\be
\delta_i^-=\vU_i-\vU_{i-1}\ ,\qquad
\delta_i^+=\vU_{i+1}-\vU_i\ .
\en
We then employ the monotonized-central (MC) limiter, for which the reconstructed slope is
\be \label{eq:MCslope}
\boldsymbol{\Delta}_i=\operatorname{minmod}\left(2\delta_i^-,\frac{\delta_i^-+\delta_i^+}{2},2\delta_i^+\right)\ .
\en
Here, the minmod function is applied componentwise to the conservative variables.\footnote{For scalar arguments, the three-argument minmod function is defined as
$\operatorname{minmod}(a,b,c)=\min(a,b,c)$ if $a,b,c>0$,
$\operatorname{minmod}(a,b,c)=\max(a,b,c)$ if $a,b,c<0$,
and $\operatorname{minmod}(a,b,c)=0$ otherwise. Thus, when the three candidate slopes have a common sign, the slope with the smallest absolute magnitude is selected; if their signs differ, the reconstructed slope is set to zero.}
The role of the limiter is to retain the centered second-order reconstruction where the solution varies smoothly, while reducing the slope near strong gradients or local extrema in order to suppress spurious numerical oscillations.

The corresponding left and right states at the interface $x_{i+1/2}$ are
\begin{align}
\vU^L_{i+1/2}&= \vU_i+\frac{1}{2}\boldsymbol{\Delta}_i\ , \\
\vU^R_{i+1/2}&= \vU_{i+1}-\frac{1}{2}\boldsymbol{\Delta}_{i+1}\ ,
\end{align}
which represent the values reconstructed at the same interface from neighboring cells $i$ and $i+1$, respectively, and are used to evaluate the numerical flux across each cell interface.

For each reconstructed interface state, $\mathcal B=\beta+\delta\mathcal B$, while $\mathcal E$ is obtained from $d=\ell_f\paren{f}\mathcal E$. On the branch connected to the Maxwell limit, the Born-Infeld and logarithmic constitutive relations yield
\begin{align}
\mathcal E_\text{BI}
 &=\frac{d\sqrt{1+\mathcal B^2}}{\sqrt{1+d^2}}\ ,\\
\mathcal E_\text{log}
 &=\frac{d\paren{2+\mathcal B^2}}
 {1+\sqrt{1+d^2\paren{2+\mathcal B^2}}}\ .
\end{align}
For the polynomial model, the corresponding value of $\mathcal E_p$ is the real root connected continuously to $\mathcal E_p=0$ at $d=0$ of
\be
\mathcal E_p^3+\paren{2-\mathcal B^2}\mathcal E_p-2d=0\ .
\en
Once the primitive variables are recovered, we evaluate the physical flux using $h=\ell_f\paren{f}\mathcal B$.

Once the left and right states at each cell interface have been reconstructed, a single numerical flux must be specified in order to determine the exchange of the conserved variables between neighboring cells. Since the characteristic structure of the NLED system depends on the local electromagnetic state, and increasingly steep gradients may develop during the evolution, it is convenient to use a robust approximate Riemann solver whose dissipation is controlled by the local propagation speeds. For this purpose, we employ the local Lax-Friedrichs, or Rusanov, flux \cite{Rusanov1962,LeVeque2002},
\begin{align}
\widehat{\vFc}_{i+1/2}=&\frac{1}{2}\left[ \vFc\paren{\vU_{i+1/2}^L}+\vFc\paren{\vU_{i+1/2}^R}\right] \label{eq:Rusanov}\\
&-\frac{a_{i+1/2}}{2}\paren{\vU_{i+1/2}^R-\vU_{i+1/2}^L}\ , \nonumber
\end{align}
where
\[
a_{i+1/2}=\max_{\alpha=L,R}\left\{\left|\lambda_+\paren{\vU_{i+1/2}^\alpha} \right|,\left| \lambda_-\paren{\vU_{i+1/2}^\alpha}\right| \right\}\ ,
\]
is the largest absolute characteristic velocity associated with either reconstructed state at the interface. Equivalently, $a_{i+1/2}$ provides a local estimate of the spectral radius of the flux Jacobian and therefore represents the largest signal speed that can propagate across the interface.

The first term in Eq. \eqref{eq:Rusanov} is the average of the physical fluxes evaluated on the two sides of the interface, whereas the second introduces a dissipative correction proportional to the jump between the reconstructed states. The strength of this numerical dissipation is controlled by $a_{i+1/2}$ and is therefore determined locally by the characteristic structure of the NLED system. Since $\lambda_{\pm}$ depend on the instantaneous electromagnetic state, the dissipative contribution adapts dynamically as the wave evolves.

The finite-volume discretization then leads to the semidiscrete system
\be \label{eq:semidiscrete_FV}
\frac{d\vU_i}{dt}=-\frac{\widehat{\vFc}_{i+1/2}-\widehat{\vFc}_{i-1/2}}{\Delta x}\ .
\en
Thus, the rate of change of the cell-averaged state is determined by the imbalance between the numerical fluxes entering and leaving the cell.

The system in Eq. \eqref{eq:semidiscrete_FV} is integrated in time using the third-order strong-stability-preserving Runge-Kutta method (SSP-RK3) \cite{Shu1988}. The timestep is adjusted dynamically according to the Courant-Friedrichs-Lewy (CFL) condition,
\be
\Delta t=C_\text{CFL}\frac{\Delta x}{\max_i a_{i+1/2}}\ ,
\en
where we set $C_{\text{CFL}}=0.4$ for all the calculations presented below. Since the characteristic velocities depend on the local electromagnetic state, the maximum propagation speed is recalculated throughout the evolution and the timestep is updated accordingly. 

Periodic boundary conditions are imposed on a computational domain sufficiently large that the propagating wave packet does not interact with its periodic image over the time interval considered. The conservative character of the scheme also allows the spatially integrated quantities associated with $d$ and $\delta\mathcal{B}$ to be monitored throughout the evolution as numerical consistency checks.

\subsection{Peak-gradient comparisons}

Tables~\ref{tab:table2} and \ref{tab:table3} give the numerical and characteristic values of $G=\max_x|\partial_x\mathcal E|$ for the two validation cases in Fig.~\ref{fig:fig2}. Both use $\beta=0.5$ and $\sigma=2$; the logarithmic and polynomial pulses have $A=0.6$ and $A=0.2$, respectively.  Both validation runs used the same spatial resolution, $\Delta x=0.05$. The logarithmic calculation used $N=1600$ cells on an 80-unit domain, whereas the polynomial calculation used $N=1200$ cells on a 60-unit domain. Agreement is closest while the profiles are smooth. Near $t_{\rm sh}$ the finite-volume method regularizes the steepening front, so its peak gradient falls below the characteristic prediction.

For the logarithmic pulse, an additional run with $N=3200$ and $\Delta x=0.025$ on an 80-unit domain provides a resolution check near the catastrophe. At $t/t_{\rm sh}^{\log}=0.75$, the higher-resolution peak gradient is $0.650703$, compared with $0.594955$ at $N=1600$ and the characteristic value $0.684261$. At $t/t_{\rm sh}^{\log}=0.90$, the corresponding values are $1.16642$, $1.09572$, and $1.69193$. Increasing the resolution moves the finite-volume values toward the characteristic prediction at both times, while a substantial difference remains close to $t_{\rm sh}$.

\begin{table}[t]
\centering
\begin{tabular}{c c c}
\hline \hline
$t/t_{\text{sh}}^{\log}$ & $G_{\text{FV}}$ & $G_{\text{char}}$ \\
\hline
$0$ & $0.181950$ & $0.181959$ \\
$0.25$ & $0.237134$ & $0.237531$ \\
$0.50$ & $0.344286$ & $0.348939$ \\
$0.75$ & $0.594955$ & $0.684261$ \\
$0.90$ & $1.09572$ & $1.69193$ \\
\hline \hline
\end{tabular}
\caption{Peak gradients for the logarithmic validation pulse in Fig.~\ref{fig:fig2}.}
\label{tab:table2}
\end{table}

\begin{table}[t]
\centering
\begin{tabular}{c c c}
\hline\hline
$t/t_{\rm sh}^{p}$ & $G_{\rm FV}$ & $G_{\rm char}$ \\
\hline
$0$    & $0.0606499$ & $0.0606531$ \\
$0.25$ & $0.076438$  & $0.0765001$ \\
$0.50$ & $0.107682$  & $0.108465$  \\
$0.75$ & $0.188211$  & $0.206238$  \\
$0.90$ & $0.327024$  & $0.502359$  \\
\hline\hline
\end{tabular}
\caption{Peak gradients for the polynomial validation pulse in Fig.~\ref{fig:fig2}.}
\label{tab:table3}
\end{table}

\FloatBarrier
\bibliography{2+1D_waves}

\end{document}